\documentclass[a4paper,UKenglish]{lipics-v2016}

\usepackage{booktabs} % For formal tables

\usepackage{amsmath,amsfonts,amssymb,bbm}
\usepackage{comment}
\usepackage{setspace}
\usepackage{cancel}
\usepackage{paralist}
\usepackage{caption}
\renewcommand{\paragraph}[1]{{\protect\vspace{8pt}\noindent\sc{#1}}}
\usepackage{multicol}
\usepackage{dsfont}
\usepackage{enumerate}
\newlength{\saveparindent}
\newlength{\saveparskip}
\newcommand{\BE}{\begin{enumerate}} \newcommand{\EE}{\end{enumerate}}
\newcommand{\BI}{\begin{itemize}} \newcommand{\EI}{\end{itemize}}
\newcommand{\BDes}{\begin{description}}\newcommand{\EDes}{\end{description}}
\newtheorem{alg}{Algorithm}
\newcommand{\BA}{\begin{alg}} \newcommand{\EA}{\end{alg}}
 \newtheorem{thm}{Theorem}[section]            % A counter for Theorems
\newcommand{\BT}{\begin{thm}} \newcommand{\ET}{\end{thm}}

\newtheorem{lem}[thm]{Lemma} % Changed counter to be as for Theorems
\newcommand{\BL}{\begin{lem}} \newcommand{\EL}{\end{lem}}

\newtheorem{clm}[thm]{Claim}
\newcommand{\BCM}{\begin{clm}} \newcommand{\ECM}{\end{clm}}

\newtheorem{techcor}[thm]{Corollary}
\newcommand{\BCo}{\begin{techcor}} \newcommand{\ECo}{\end{techcor}}

\newtheorem{Conc}[thm]{Conclusion}
\newcommand{\BCONC}{\begin{Conc}} \newcommand{\ECONC}{\end{Conc}}

\newtheorem{Obs}[thm]{Observation}
\newcommand{\BOBS}{\begin{Obs}} \newcommand{\EOBS}{\end{Obs}}

\newtheorem{Exmp}[thm]{Example}
\newcommand{\BEXM}{\begin{Exmp}} \newcommand{\EXMP}{\end{Exmp}}

\newtheorem{cor} [thm] {Corollary}      % counter AS FOR Theorems
\newcommand{\BC}{\begin{cor}} \newcommand{\EC}{\end{cor}}
\newtheorem{prop}[thm]{Proposition}     % A counter AS FOR Thms
\newcommand{\BP}{\begin{prop}} \newcommand {\EP}{\end{prop}}
\newtheorem{conj} {Conjecture}      % counter AS FOR Theorems
\newcommand{\BCJ}{\begin{conj}} \newcommand{\ECJ}{\end{conj}}
\newtheorem{claim} {Claim}      % counter AS FOR Theorems
\newcommand{\BCL}{\begin{claim}} \newcommand{\ECL}{\end{claim}}

\newtheorem{fact}[thm]{Fact}
\newcommand{\BF}{\begin{fact}} \newcommand{\EF}{\end{fact}}
\newtheorem{assumption}[thm]{Assumption}
\newcommand{\BAs}{\begin{assumption}} \newcommand{\EAs}{\end{assumption}}
\newtheorem{defn}{Definition}[section]         % A counter for Definition
\newcommand{\BD}{\begin{defn}} \newcommand{\ED}{\end{defn}}
\def\FullBox{\hbox{\vrule width 8pt height 8pt depth 0pt}}
\newcommand{\QED}{\;\;\;\FullBox}
\newenvironment{Proof}{\noindent{\bf Proof:~~}}{\hfill\QED}
\newcommand{\BPF}{\begin{Proof}} \newcommand {\EPF}{\end{Proof}}
\newenvironment{proofof}[1]{\noindent{\bf Proof of {#1}:~~}}{\(\hfill\QED\)}
\newcommand{\BPFOF}{\begin{proofof}} \newcommand {\EPFOF}{\end{proofof}}

\newenvironment{smallproof}{\noindent{\bf Proof sketch:~~}}{\(\QED\)}
\newcommand{\bpf}{\begin{smallproof}} \newcommand{\epf}{\end{smallproof}}
\newcommand{\BEQ}{\begin{equation}} \newcommand{\EEQ}{\end{equation}}
\newcommand{\BEQN}{\begin{eqnarray}}\newcommand{\EEQN}{\end{eqnarray}}
\renewcommand{\Pr}{{\rm Pr}}

\newcommand{\pr}{{\mbox{\bf\rm Pr}}}

\newcommand{\eps}{\epsilon}

\newcommand*{\rom}[1]{\expandafter\@slowromancap\romannumeral #1@}

\newcommand{\E}{{\rm E}}

\newcommand{\argmax}{{\rm argmax}}

\usepackage{mathtools}

\DeclarePairedDelimiter\abs{\lvert}{\rvert}%
\usepackage{thm-restate}

\newcommand{\MyFrame}[1]{\noindent \framebox[\textwidth]{ \begin{minipage}{0.97\textwidth} #1 \end{minipage}}}%

\newcommand{\qedsymb}{\hfill{\rule{2mm}{2mm}}}
\newcommand{\rseq}{\mathcal{R}_{\mathrm{seq}}}
\newcommand{\rsim}{\mathcal{R}_{\mathrm{sim}}}
\newcommand{\p}{\mathbf{p}}

\newcommand{\TT}{\mathbf{T}}
\newcommand{\tr}{\mathbf{T}_{\F}}
\newcommand{\rmax}{\overline{\mathcal{R}}_{\mathrm{sim}}}
\newcommand{\rmin}{\underline{\mathcal{R}}_{\mathrm{sim}}}
\newcommand{\rbestsim}{\mathcal{R}^*_{\mathrm{sim}}}
\newcommand{\rbwsim}{\mathcal{\underline{R}}^*_{\mathrm{sim}}}
\newcommand{\rbestseq}{\mathcal{R}^*_{\mathrm{seq}}}
\newcommand{\rbwseq}{\mathcal{\underline{R}}^*_{\mathrm{seq}}}

\newcommand{\revpubseqmult}{\mathcal{R}^k}
\newcommand{\rev}{\mathcal{R}}

\newcommand{\F}{\mathcal{F}}
\newcommand{\RR}{\mathcal{R}}
\newcommand{\PI}{\mathcal{PI}}

\newcommand{\w}{\mathbf{w}}
\newcommand{\piprice}{p^{\mathcal{PI}}}
\newcommand{\piprices}{\p^{\mathcal{PI}}}
\newcommand{\feas}{\mathcal{U}}

\usepackage{units}

\usepackage{color-edits}
\addauthor{ae}{green}
\addauthor{mf}{blue}
\addauthor{te}{purple}

\usepackage{microtype}%if unwanted, comment out or use option "draft"

\title{Pricing Social Goods} %\footnote{This work was partially supported by someone.}}
\titlerunning{Pricing Social Goods} %optional, in case that the title is too long; the running title should fit into the top page column

\author[1]{Alon Eden}
\author[1]{Tomer Ezra}
\author[2]{Michal Feldman}
\affil[1]{Tel Aviv University, Israel\\
	\texttt{\{alonarde,tomer.ezra\}@gmail.com}}
\affil[2]{Tel Aviv University and Microsoft Research, Israel\\
	\texttt{michal.feldman@cs.tau.ac.il}}
\authorrunning{A. Eden, T. Ezra and M. Feldman} %mandatory. First: Use abbreviated first/middle names. Second (only in severe cases): Use first author plus 'et. al.'

\begin{document}
	
	\maketitle
	
\begin{abstract}
	Social goods are goods that grant value not only to their owners but also to the owners' surroundings, be it their families, friends or office mates.
	The benefit a non-owner derives from the good is affected by many factors, including the type of the good, its availability, and the social status of the non-owner.
	Depending on the magnitude of the benefit and on the price of the good, a potential buyer might stay away from purchasing the good, hoping to free ride on others' purchases.
	A revenue-maximizing seller who sells social goods must take these considerations into account when setting prices for the good.
	The literature on optimal pricing has advanced considerably over the last decade, but little is known about optimal pricing schemes for selling social goods.
	In this paper, we conduct a systematic study of revenue-maximizing pricing schemes for social goods: 
	we introduce a Bayesian model for this scenario, and devise nearly-optimal pricing schemes for various types of externalities, both for simultaneous sales and for sequential sales.
\end{abstract}

	\section{Introduction}
	Many goods exhibit a positive externality not only on their owner, but also on other parties. For instance, a coffee machine purchased by an employee benefits all of her office mates, and essentially reduces the probability of another coffee machine to be purchased. Examples of these kinds of goods are abundant: A high-schooler who has many friends with cars that can drive him around might be less tempted to buy a new car. A reputable store might draw  large customer traffic and benefit other stores in the shopping mall. Therefore, an aggressive advertising campaign carried out by such a store might reduce the likelihood of another store running a campaign in parallel.
In all of these scenarios the externalities depend on the type of good, on the social status of the party with whom the good is shared, and on the set of parties who own the good. In the coffee machine example, the machine is typically used by all the individuals sharing the office space. In the shopping mall, some types of stores (\textit{e.g.}, fast food restaurants) might benefit from any traffic in the shopping mall, whereas more specialized stores may benefit from ad campaigns that draw costumers interested in a similar kind of product (\textit{e.g.}, \texttt{Staples} may attract costumers similar to those interested in \texttt{Office Depot} products). The benefit of a high school student depends on his social status and on the set of friends who own a car.

Because of the abundance of goods that exhibit externalities similar to the ones in the examples above, their study is of great applicability. We term these goods \textit{social goods}. When selling social goods, a seller must take into account the types of buyers in the market and the benefit they derive from other sets of buyers purchasing the good. Our main goal is to study how to sell goods in a way that approximately maximizes the seller's revenue in the presence of externalities.

To study this problem, we consider a setting with a single type of good, of unlimited supply, and a set of $n$ agents; each agent $i\in [n]$ has a non-negative valuation $v_i$ for purchasing the good, drawn independently from a distribution $F_i$. We denote the product distribution by $\F= \times_{i\in [n]} F_i$.  %An agent $i$ who purchases the good derives value $v_i$ from it.
Unless stated otherwise, we assume the $F_i$'s  are regular.\footnote{This means that the virtual valuation function $\phi(v)=v-\frac{1-F(v)}{f(v)}$ is non-decreasing. See Appendix A in the full version.}

If an agent does not purchase the good, but the good is purchased by others, then this agent derives only a fraction of her value, depending on the set of agents and the type of externality the good exhibits on the agent. This type of externality is captured in our model by an \textit{externality function} $x_i:2^{[n]}\rightarrow [0,1]$, where $x_i(S)$ denotes the fraction of $v_i$ an agent $i$ derives when the good is purchased by the set of agents $S$. We assume that $x_i$ is publicly known (as it captures the agent's externalities), monotonically non-decreasing and normalized; i.e., $x_i(\emptyset)=0$, for every $T \subseteq S$, $x_i(T) \leq x_i(S)$, and $x_i(S)=1$ whenever $i\in S$.
%%For each scenario we consider, we instantiate $x_i$ to capture a different social aspect of goods.
We consider three structures of the function $x_i$, corresponding to three types of externalities of social goods.
\begin{enumerate}[(a)]
	\item \textit{Full externalities} (commonly known as ``public goods"): in this scenario all agents derive their entire value if the good is purchased by any agent. Therefore, $x_i(S)=1$ if and only if $S\neq \emptyset$. This model captures goods that are non-excludable, such as a coffee machine in a shared office. A special case of this scenario, where valuations are independently and identically distributed, has been studied in \cite{feldman2013pricing}.
	\item  \textit{Status-based externalities}: in this scenario, agent $i$'s ``social status" is captured by some \textit{discount factor} $w_i\in[0,1]$, which corresponds to the fraction of the value an agent $i$ derives from a good when purchased by another party. That is,
	\begin{eqnarray}
	x_i(S)=\begin{cases}
	1 & \quad i\in S,\\
	w_i & \quad i\notin S \mbox{ and }S\neq \emptyset,\\
	0 & \quad \mbox{otherwise}.
	\end{cases}
	\end{eqnarray}
	This model captures settings that exhibit asymmetry with respect to the benefit different agents derive from goods they do not own (\textit{e.g.}, a fast food restaurant or a popular high-school student in the above examples). %This captures also cases where the externalities are not full, and the agents \textit{are} symmetric.
	
	\item \textit{Availability-based externalities}: in this scenario, the availability of a good increases as more agents purchase a good, and therefore, an agent derives a larger fraction of her value as more agents purchase a good. This is captured by the following externality function.
	\begin{eqnarray}
	x_i(S)=\begin{cases}
	1 & \quad i\in S,\\
	w(\abs{S}) & \quad i\notin S.
	\end{cases}
	\end{eqnarray}
	Here, $w:\{0,\ldots,n-1\}\rightarrow [0,1]$ is a monotonically non-decreasing function with $w(0)=0$.
	Examples of such scenarios include objects that are often shared by neighbors (e.g., snow blowers, lawn mowers), office supplies, etc.
\end{enumerate}
Notice that the full externalities scenario is a special case of both the social-status (where $w_i=1$ for every $i$) and the availability (where $w(k)=1$ for every $k>0$) models. 	
%[[describe the three models: full externalities, social status-based externalities, availability-based externalities]]

Our focus is on posted-price mechanisms, which exhibit many desired properties: they are simple, distributed, straightforward, and strategyproof.
Our goal is to maximize the revenue extracted by the seller. We distinguish between discriminatory and non-discriminatory prices. Naturally, using discriminatory prices can often lead to higher revenue for the seller \cite{myerson1981optimal,HR09}. Price discrimination is commonly used in the US \cite{hannak2014measuring}, but user studies reveal that many users believe that this practice is illegal, and consider these acts to be an invasion of privacy \cite{calo2013digital}. Therefore, offering non-discriminatory prices may be critical for maintaining the seller's reputation. We show scenarios in which setting the same price for all users produces (almost) as much revenue as engaging in price discrimination.

We consider two natural sale models: (a) a \textit{simultaneous sale}, where the seller simultaneously sets take-it-or-leave-it prices for all agents, after which agents play a simultaneous Bayesian game, and each agent decides whether or not to buy at the price offered to her; and (b) a \textit{sequential sale}, in which the agents arrive sequentially, and each one is offered a take-it-or-leave-it price upon arrival. In this case, the price and the agent's decision may depend on the set of agents that purchased the good before the arrival of the current agent.
We distinguish between {\em adaptive} and {\em non-adaptive} pricing schemes, which differ in whether the price can depend upon the set of agents who purchased the good prior to the agent's arrival.

In both simultaneous and sequential sales, assuming that agent $i$ is offered a take-it-or-leave-it price $p_i$, and that the good is eventually purchased by a set $S\subseteq [n]$ of agents, the utility of agent $i$ is:
\begin{eqnarray}
u_i(S,p_i)=\begin{cases} v_i-p_i \quad & \mbox{if } i\in S,\\
v_i\cdot x_i(S) \quad & \mbox{if }i\notin S.\end{cases}\label{eq:utility}
\end{eqnarray}

As shown in Section \ref{sec:model}, a set of prices induces equilibria of the game (multiple equilibria in the simultaneous model, and a single one in the sequential model). Every equilibrium is characterized by a set of {\em threshold} strategies for the agents, where an agent buys the good if and only if her value exceeds the threshold.

\subsection{Our contribution}

We provide results for the three models described above (See Fig. \ref{fig:results-table}).

\paragraph{\textbf{(a) Full externalities.}}
%We devise nearly optimal pricing schemes for both the simultaneous and sequential models.

\vspace{0.1in}

\noindent {\bf Theorem (informal):}
There exist poly-time algorithms for computing pricing schemes for settings with full externalities that give a constant factor approximation to the optimal pricing scheme, for both simultaneous and sequential sales. Moreover, this result can be achieved using non-discriminatory prices, despite asymmetry among buyers.

\vspace{0.15in}

To derive this result, we first analyze the equilibria in simultaneous and sequential models.
We show a surprising equivalence between the revenue attainable in the best equilibrium at simultaneous and sequential sales, albeit induced by different prices.
A corollary of this equivalence is that the optimal attainable revenue at a sequential sale does not depend on the order of agents.
%
%We continue to investigate near optimal posted price mechanisms for this scenario.
Furthermore, we observe that in both simultaneous and sequential sales, the revenue attainable is upper bounded by the optimal revenue from selling a single {\em private} good (i.e., a good that grants value only to their owners)\footnote{A similar argument was used in \cite{feldman2013pricing} for the special case of simultaneous sales where valuations are identically distributed.}.

We proceed as follows.
For simultaneous sales, we establish a method for transforming prices for the sale of a single private good \textit{in expectation} into prices for selling public goods, which preserve the revenue up to a constant factor in every equilibrium. Since selling a single good in expectation yields at least as much revenue as selling a single good deterministically, this implies a near-optimal pricing scheme for simultaneous sales of public goods.

For sequential sales, we use the theory of prophet inequalities.
Consider prices that induce thresholds that are equal to the prices that emerge from the prophet inequalities.
We show that such prices obtain at least half of the revenue that is obtained from the prophet inequalities prices in the private good model.
We use this connection to obtain a pricing scheme that gives 4-approximation to the revenue of the optimal sequential sale of public goods.

Finally, we show how to compute nearly-optimal \textit{non-discriminatory} prices, even for asymmetric agents, in both the simultaneous and sequential models.

\paragraph{\textbf{(b) Status-based externalities.}}

\vspace{0.1in}

\noindent {\bf Theorem (informal):} There exist poly-time algorithms for computing pricing schemes for settings with status-based externalities that give a constant factor approximation to the optimal pricing scheme, for both simultaneous and sequential sales.\footnote{We note that no non-discriminatory prices can achieve a constant approximation in this model. Indeed, the case of private digital goods is a special case of this model, with $w_i=0$ for every $i$.}

\vspace{0.15in}

For sequential sales, we devise a {\em non-adaptive} pricing scheme, while the benchmark is the optimal adaptive pricing scheme. To obtain this result, we first show that a seller who is restricted to set only two prices per agent can extract as much revenue as one who can present exponentially many prices.
We then show that the optimal revenue in this simpler case can be decomposed into two components: a private component (monotonically decreasing in the agents' discount  factors) and a public component (monotonically increasing in the discount factors).
The private component can be approximated by simulating $n$ private sales, setting thresholds equal to the monopoly prices.
The public component can be approximated by similar techniques to the ones introduced for public goods.
Therefore, the better of the two mechanisms extracts a constant fraction of the optimal revenue.
A similar decomposition technique is established for the case of simultaneous sales.
Our result for the sequential case is essentially a reduction: given prices that yield a $c$-approximation for the optimal sequential sale in the full externalities model, one can find prices that $(c+2)$-approximate the optimal sequential sale in the status-based externalities model.

\paragraph{\textbf{(c) Availability-based externalities.}}

\vspace{0.1in}

\noindent {\bf Theorem (informal):} There exists a poly-time algorithm for computing a pricing scheme for sequential sales with availability-based externalities, that gives a logarithmic factor approximation (with regard to the number of buyers) to the optimal pricing scheme.

\vspace{0.15in}

In this case, both the pricing scheme and the benchmark set a pricing function for each agent, which depends on the number of agents who have purchased the good before the arrival of the agent.
To obtain this result, we decompose the revenue into $n$ components. Component $k=1,\ldots, n$ is upper bounded by the optimal revenue obtainable by selling $k$ identical private goods, scaled by $w(k)-w(k-1)$.
%to agents who arrive in the same order and have the same priors multiplied by the contribution of the $k$th agent to purchase to the externality function.
We then partition the components into buckets, and compute prices based on the sequential posted pricing scheme developed by  Chawla et al. \cite{chawla2010multi} for selling private goods.

\paragraph{\textbf{General externalities.}}
Given the near-optimal pricing schemes above, one may be tempted to infer that every social goods scenario is amenable to a near-optimal pricing scheme.
We complement our positive results with the following hardness result, refuting this hope.
We consider a natural family of social goods proposed by Feldman et al. \cite{feldman2013pricing}: network-based externalities. In this model, externalities are represented by a graph, and an agent derives her entire value when a neighboring agent buys a good. We show that there is no poly-time algorithm to compute prices that give a non-trivial approximation to the optimal posted-price mechanism. This negative result holds for both the simultaneous and sequential models. We show that even in very restricted cases %the simple case in which
(i.e., where agents' valuations are independently and uniformly distributed on $[0,1]$ in the simultaneous case, and agents' valuations are fixed in the sequential case), it is NP-hard to find prices that approximate the optimal posted-price mechanism to within a factor of $n^{1-\epsilon}$.
A $\Theta(n)$ approximation can be trivially achieved by offering the good only to the agent maximizing the monopolist revenue.
We note that this negative result rules out other natural externality structures.\footnote{Some examples include: (a) for every pair of agents $i,j$, agent $i$ can borrow the good from agent $j$ with some probability $w_{ij}$. Thus, $x_i(S) = 1-\prod_{j\in S, j\neq i}(1-w_{ij})$; and (b) for every pair of agents $i,j$, $x_i(S) = \max_{j\in S, j\neq i}w_{ij}$.}

\paragraph{\textbf{Irregular distributions.}}
Although our results are stated and proved for regular distributions, %mostly for ease of presentation,
some of our results extend to irregular distributions.
Namely, we establish near optimal pricing schemes for sequential and simultaneous sales under full externalities and status-based externalities. 
%, as discussed in Appendix~\ref{app:irregular}. 
The results of non-discriminatory prices do not extend to irregular distributions since the anonymous pricing devised in \cite{DBLP:journals/corr/AlaeiHNPY15} do not perform well for irregular distributions. (As shown in \cite{DBLP:journals/corr/AlaeiHNPY15}, there exist irregular distributions that do not admit any anonymous prices that give constant approximation.)

		\begin{figure}[h]				
	\setlength{\fboxrule}{0.5 pt}
	\noindent \fbox{\noindent\makebox[1\textwidth][c]
		{\begin{minipage}{1\textwidth}	
				
				\centering
				\scalebox{0.75}{
				\begin{tabular}{|c|c|ll|ll|ll|ll}
					\cline{3-10} \multicolumn{2}{l|}{ }  & \multicolumn{4}{c|}{Simultaneous} & \multicolumn{4}{c|}{Sequential}\\
					\cline{3-10} \multicolumn{2}{l|}{ }  & \multicolumn{2}{c|}{discriminatory} & \multicolumn{2}{c|}{non-discriminatory} & \multicolumn{2}{c|}{discriminatory} & \multicolumn{2}{c|}{non-discriminatory}\\
					\hline Full (Public goods) & i.i.d. & $\geq 4/e$ & Thm. \ref{thm:lb} & $4$ & Thm. \ref{thm:iid}  & \multicolumn{2}{c|}{-} & $4$ & \multicolumn{1}{c|}{Thm. \ref{thm:iid}} \\
					\cline{2-10}   & non i.i.d. & $5.83$ & Thm. \ref{thm:exante_reduction} & $4e$ & Cor. \ref{cor:nondisc}  & $4$ & Thm. \ref{thm:canonical-rev} & $4e$ & \multicolumn{1}{c|}{Cor. \ref{cor:nondisc}} \\
					\hline \multicolumn{2}{|c|}{Status-based} & $6.83$ & Thm. \ref{thm:semi_sim_main} & $\Omega(\log n)$ & App. \ref{app:inapprox_homog} & 6 & Cor. \ref{cor:status_seq_approx} & $\Omega(\log n)$ & \multicolumn{1}{c|}{App. \ref{app:inapprox_homog}} \\
					\hline \multicolumn{2}{|c|}{availability-based} &  \multicolumn{2}{c|}{-}& \multicolumn{2}{c|}{-} & $O(\log n)$ & Thm. \ref{thm:grad_main} & \multicolumn{2}{c|}{-}\\
					\hline \multicolumn{2}{|c|}{network-based} & $\Omega\left(n^{1-\eps}\right)$ & Thm. \ref{thm:hard_sim} & \multicolumn{2}{c|}{-} & $\Omega\left(n^{1-\eps}\right)$ & \multicolumn{1}{c|}{Thm. \ref{thm:hard_seq}} &  \multicolumn{2}{c|}{-} \\ 
					\hline
				\end{tabular} 
				}
				\caption{Summary of our results for pricing mechanism for social goods.
					The columns correspond to sale models, whereas the rows correspond to types of externalities. The rows are further divided to sales using discriminatory and non-discriminatory prices. 
				}					
				
				\label{fig:results-table}
				
	\end{minipage}}} \par\setlength{\fboxrule}{0.2pt}								
	
\end{figure}
\begin{comment}
\begin{table}[h]
	\begin{center}
		\begin{tabular}{|c|c|c|c|}
			\hline
			Externalities / Sale & Simultaneous  & Sequential & 
			
			 Full & Status-based & Availability-based & Graph-based  \\ \hline
			& $5.83$ \ref{thm:exante_reduction}& $6.83$ \ref{thm:semi_sim_main} & & Hardness result \ref{thm:hard_sim} \\ \hline
			 & $4$  \ref{thm:canonical-rev} & $6$ \ref{thm:semi_seq_main} &  $\Theta(logn)$ \ref{thm:grad_main}  & Hardness result \ref{thm:hard_seq}  \\ \hline
			Uniform prices & $4e$ tbd addlabel  & appendix G &  &  \\ \hline
			
		\end{tabular}
		\caption{}\label{t:sum}
	\end{center}
\end{table}
\end{comment}
%As opposed goods who exhibit full externalities, we cannot bound the optimal revenue obtained by a sale of goods who exhibit social status-based externalities using the optimal revenue obtained by selling a single private good, as when $w_i=0$ for every agent $i$ the setting is equivalent to the sale of digital private goods. Some technical work allows us to decompose the revenue obtainable to two components

\subsection{Related work}\label{sec:related}
The most famous and well studied instance of social goods is public goods, when all agents derive their full value whenever a good is purchased.
The study of public goods was initiated by Samuelson \cite{samuelson1954pure}, who observed that private provisioning of public goods is not necessarily efficient; see also \cite{mas1995microeconomic} for an overview.
%Public goods have also been studied in the context of market equilibria. Bergstrom et al.
%\cite{bergstrom1986private} and Allouch \cite{allouch2015private} studied the existence of market equilibria and their properties in settings in which agents have initial endowments of both public and private goods.

%Locally public goods were introduced by \cite{bramoulle2007public} and \cite{bramoulle2014strategic}, who modeled the externality structure using an underlying graph.
%In their model, agents decide on the level of effort they exert in a joint project, and their utilities grow as the level of effort exerted by the agents and their neighbors increases.

%Similarly to our setting,
%\cite{candogan2012optimal} considered a monopolist who sets prices for agents who exhibit positive externalities, captured by a network structure. In contrast to our Bayesian setting, they consider the case of full information. In addition, they consider a divisible good while we consider indivisible goods, and their utility function is of a different form.

The closest work to ours is that of Feldman et al. \cite{feldman2013pricing}.
For their positive results, they consider a special case of our full externalities model --- in their model agents arrive simultaneously with valuations that are drawn independently and identically from a known distribution.
Our work extends this work in several dimensions.
First, we consider more realistic forms of externalities that go beyond public goods.
Second, we consider settings where agent valuations are drawn from non-identical distributions.
Third, we provide results for settings where agents arrive either sequentially or simultaneously. 
Finally, some of our results extend to irregular distributions.
%We also consider more elaborative scenarios
%We also extend the case of perfectly public goods, considered in \cite{feldman2013pricing}, to the case of semi-public goods.

A line of work similar in flavor to ours, yet inherently different, is that of revenue maximization in the presence of positive externalities \cite{ahmadipouranari2013equilibrium,haghpanah2013optimal,hartline2008optimal,akhlaghpour2010optimal,candogan2010optimal}. In this line of work, an agent's value for the good increases as more agents purchase the goods, but only if the agent purchased the good as well. Therefore, an agent is more likely to purchase the good as more agents purchase it. This is in stark contrast to our setting, where agents are less inclined to buy a good as more agents do.

Finally, there is a rich body of literature on the design of posted price mechanisms for the sale of private goods (where agents do not derive value from goods they do not own).
See Chapter 4 in \cite{hartlineMDnA} for a textbook treatment.
A sample of the work can be found in \cite{chawla2010multi,DBLP:journals/corr/AlaeiHNPY15, kleinberg2012matroid, feldman2015combinatorial, chawla2007algorithmic}.
An overview of some results that are directly referred to in this work is given in Appendix \ref{app:singleparam}. %\ref{section:environments} and \ref{section:posted_prices}. 

\subsection{Organization}
Model and preliminaries appear in Section \ref{sec:model}. Our results for the full externalities model appear in Section \ref{sec:clique}. %, where in Section \ref{sec:public_equi} we characterize the equilibrium in both sale models, and show the equivalence of revenue in both models, in Sections \ref{sec:public_sim} and \ref{sec_seq}  we describe the near optimal sequential and simultaneous sales, and in Section \ref{sec:public_nondisc} we show how to achieve a near optimal sale using non-discriminatory prices. 
In Section \ref{sec:semi-public} we present our results for goods that exhibit status-based externalities for sequential (Section \ref{sec:fading_seq}) and simultaneous (Section \ref{sec:fading_sim}) sales. Our results for the availability-based externalities model appear in Section \ref{sec:semi-gradual}, and our hardness results appear in Section \ref{sec:obsevations}. We discuss how to extend our results to irregular distributions in Appendix \ref{app:irregular}.

	\section{Models and preliminaries}\label{sec:model}
	\begin{comment}
Our model of social goods consists of $n$ potential buyers, termed \textit{agents}. Each agent $i\in [n]$ has some non-negative valuations $v_i$, derived independently from a distribution $F_i$. The product distribution is denoted by $\F =\prod_{i\in [n]}F_i$. The social aspect of agent $i$ is captured by a function $x_i:2^{[n]}\rightarrow [0,1]$ which indicates what fraction of the agent's value she derives when a subset of the agents purchase the good, that is, if a set $S\subseteq [n]$ purchase a good, agent $i$ value from this outcome is $x_i(S)\cdot v_i$. We assume $x_i$ is monotonically non-decreasing and normalize $x_i(\emptyset)=0$ and $x_i(S)=1$ whenever $i\in S$. In each of the following sections, we instantiate $x_i$ to capture a different social aspect of goods.
\end{comment}

	%$$u_i(S,p_i)=v_i\cdot x_i(S)-p_i\cdot \mathbbm{1}\left[i\in S \right]$$
	%We assume the agents' valuations are sampled from a product distribution $\F=\prod_{i\in [n]}F_i$. In our model, the seller posts a price vector %$\mathbf{p}$ with the goal of maximizing the revenue from selling the goods to the agents.
%All the missing proofs of  this section are deferred to Appendix \ref{app:missing_model}.	
	\paragraph{\textbf{Simultaneous sales model.}}
We view a simultaneous sale game as the following two-stage game. First, the seller posts a price vector $\mathbf{p}=(p_1,\ldots, p_n)$ to the agents (agent $i$ is offered to purchase an item at price $p_i$). Subsequently, the agents play a simultaneous Bayesian game. In this model, we assume that the probability distribution of every agent is atomless.\footnote{Meaning that for every $q$ there exists $p$ for which $F_i(p)=q$.}
	
	Agents wish to maximize their expected utility. Given a price $p_i$, agent $i$ buys the good if her expected utility from buying, $v_i-p_i$, exceeds the utility from not buying, $v_i\cdot\E_{S \not\owns i}[x_i(S)]$ (where $\E_{S \not\owns i}$ is shorthand for $\E_{S: i \not\in S}$). %$v_i\cdot\E_{S \subseteq [n] \setminus \{i\}}[x_i(S)].$ % \left(1-\prod_{j\in N(i)}\left[j\mbox{ does not buy}\right]\right)$. 
  Therefore, an agent buys if and only if $v_i\geq \frac{p_i}{1-\E_{S \not\owns i}[x_i(S)]}=: T_i.$
	The strategy of every agent $i$ is therefore defined by a threshold $T_i$. Denote by $\mathbf{T} =\left( T_1,\ldots,T_n\right)$ a strategy profile, given by a vector of thresholds.
A strategy profile $\mathbf{T}$ induces a probability distribution over the set $S$ of agents that purchase the good; denote this distribution by $\mu_\TT$, %.For every set $S$, the probability that the good is purchased by $S$ is $\prod_{j\notin S} F_j(T_j)\cdot\prod_{j\in S} (1-F_j(T_j))$.
and the distribution $\mu_\TT$ conditioned on $i$ not being in the set of purchasing agents by $\mu_\TT^{-i}$.
	A Nash equilibrium is characterized by a threshold vector $\mathbf{T}$ such that:
	\begin{eqnarray}
		T_i=\frac{p_i}{1-\E_{S \sim \mu_\TT^{-i}}[x_i(S)]}\quad\forall i\in[n].\label{eq:eq_condition}
	\end{eqnarray}
	
The following theorem establishes the existence of Nash equilibria via a fixed point argument.
%Its proof follows a fixed-point argument and is deferred to Appendix \ref{app:missing_model}.

	\BT\label{thm:eq_existance}
		In the simultaneous model, for any set of externality functions $\{x_i\}_{i\in [n]}$, for any set of atomless distributions $\mathcal{F}$, and for any price vector $\mathbf{p}$, there exists an equilibrium $\mathbf{T}$. % (a vector that satisfies Eq. (\ref{eq:eq_condition})).
	\ET	
	\BPF
	Given a vector of probabilities $\mathbf{q}=(q_1,\ldots, q_n)\in [0,1]^n$ we define $\eta_{\mathbf{q}}$ to be the distribution over subsets of agents, where every agent $i$ is in the subset independently with probability $1-q_i$. We define $\eta^{-i}_{\mathbf{q}}$ as the distribution conditioned on agent $i$ not being in the sampled set.
	%\textbf{Proof of Theorem \ref{thm:eq_existance}}:
	To simplify the proof, an expression with $0$ in the denominator is regarded as having value $\infty$, and $F_i$ is extended to satisfy $F_i(\infty)=1$.
	%Let $B=[0,1]^n$.
	%\quad \forall i. B_i=\left[F_i(p_i), F_i\left(\frac{p_i}{1-\E_{S \sim \mu_\p^i}[x_i(S)]}\right)\right],$$
	%and let $$\Phi(q)=\Phi_1(\TT)\times \Phi_2(\TT)\times \ldots \times \Phi_n(\TT).$$
	Given a vector of probabilities $\mathbf{q}$ we define a function $$\Phi(\mathbf{q})=\Phi_1(\mathbf{q})\times \Phi_2(\mathbf{q})\times \ldots \times \Phi_n(\mathbf{q}),\mbox{ where } \Phi_i(\mathbf{q}\in [0,1]^n)=F_i\bigg(\frac{p_i}{1-\E_{S \sim \eta_\mathbf{q}^{-i}}[x_i(S)]}\bigg).$$ % is the best-response function of agent $i$.
	$\Phi(\mathbf{q})$ is a function from $[0,1]^n$ to $[0,1]^n$. Since the $F_i$'s are atomless, $\Phi$ is a continuous function from a compact set to itself, and Brouwer's Fixed Point Theorem applies. Thus, there exists $\mathbf{q}\in[0,1]^n$ for which $\Phi(\mathbf{q})=\mathbf{q}$.
	Given a fixed point $\mathbf{q}$, setting $T_i:= \frac{p_i}{1-\E_{S \sim \eta_\mathbf{q}^{-i}}[x_i(S)]}$ yields an equilibrium. This is true since for every $i$, $q_i=F_i(T_i)$, and therefore, $\eta_\mathbf{q}^{-i}=\mu_\mathbf{\TT}^{-i}$, which implies that the equilibrium condition defined in Eq. $(\ref{eq:eq_condition})$ holds.
	\EPF

One of the challenges in our model stems from the fact that a single price vector may induce multiple equilibria. Consider the simple setting of a single public good and two agents, Alice and Bob, where $F_{\mathrm{Alice}}$ and $F_{\mathrm{Bob}}$ are both uniform on $[0,1]$, and the seller sets a non discriminatory price of $1/2$. Applying the equilibrium condition in Eq. $(\ref{eq:eq_condition})$, we get that every tuple $(T_{\mathrm{Alice}},T_{\mathrm{Bob}})\in[0,1]^2$ satisfying $T_{\mathrm{Alice}}\cdot T_{\mathrm{Bob}}=1/2$ forms an equilibrium strategy.\footnote{For a comprehensive discussion regarding the equilibrium condition in the public goods model, see Eq.(\ref{eq:global_sim_eq}) in Section \ref{sec:clique}.} Therefore, in this case, there is a continuum of equilibria. It is not hard to see, however, that a set of thresholds $\TT$ can be the consequence of only a single price vector, which can be derived via Eq. $(\ref{eq:eq_condition})$.
This is cast in the following observation:
	\BOBS\label{obs:sim_single_price}
		In the simultaneous model, a given price vector can induce multiple equilibria, but any given equilibrium $\TT$ can be induced by a single price vector $\p$. %, which can be calculated using Equation (\ref{eq:eq_condition}).
	\EOBS
	
Let $Eq(\F,\p)$ denote the set of equilibria induced by a price vector $\p$, given a product distribution $\F$.
For a given price vector $\p$ and an equilibrium $\TT\in Eq(\F,\p)$, let $\rsim(\F,\p,\TT)=\sum_i p_i\cdot (1-F_i(T_i))$ denote the seller's expected revenue. % revenue is given by
Given a price vector $\p$, we define
$$\rmax(\F,\p)=\max_{\TT\in  Eq(\F,\p)}\rsim(\F,\p,\TT)\ \ \ \ \ \mbox{and}\ \ \ \ \ \rmin(\F,\p)=\min_{\TT\in  Eq(\F,\p)}\rsim(\F,\p,\TT)$$ to be the revenue obtained in the respective best and worst equilibrium induced by $\p$. We refer to these revenues as the {\em optimistic} and {\em pessimistic} revenues, respectively.

The strongest approximation results one can hope for are ones that consider the pessimistic revenue obtained by our pricing scheme against an optimistic benchmark.
This is exactly the approach we take.
In particular, our benchmark is the revenue obtained by the best pricing, assuming the best equilibrium induced by every pricing. We denote the benchmark by $\rbestsim(\F)=\max_{\p^*}\rmax(\F,\p^*)$.
The performance of a price vector $\p$ is measured by the worst equilibrium induced by $\p$; i.e., $\rmin(\F,\p)$.
Our goal is to calculate a price vector $\p$ that minimizes the ratio between the former and the latter expressions.

	\paragraph{\textbf{Sequential sales model.}}
In the sequential sales model, $n$ agents arrive one by one according to an order $\sigma:[n]\rightarrow [n]$, where agent $i$ is the $\sigma(i)$th agent to arrive. For ease of notation, we assume that agent $i$ is the $i$th agent to arrive, unless explicitly stated otherwise.
	In sequential sales, the price set by the seller for agent $i$ can depend on the set of agents who have purchased the good prior to agent $i$'s arrival. Thus, it can be viewed as a function $p_i:2^{[i-1]}\rightarrow \mathbb{R}^+$.\footnote{Indeed, there are cases where the seller can gain higher revenue by setting such prices (an explicit example for availability-based externalities is given in Appendix \ref{app:example-menu}).}
The subgame perfect equilibrium in this auction is unique and can be found by a (possibly exponential) backward induction. An agent who receives a price buys if and only if her utility from buying exceeds her expected derived value from not buying conditioned on the set of agents that purchased the good prior to her arrival. Of course, this might impose a different threshold for every scenario which might lead to an exponential strategy space for the agents and an exponential time to compute each threshold in the strategy of an agent. As we discuss in the following sections, we devise pricing schemes in which the seller has a simple nearly optimal pricing scheme which leads to a simple strategy space and a poly-time threshold computation.

	\section{Pricing goods with full externalities (public goods)}\label{sec:clique}
	
In Section \ref{sec:public_equi} we characterize the equilibrium in both simultaneous and sequential sales models, and show an equivalence between the optimal posted prices revenue in both models. In Sections \ref{sec:public_sim} and \ref{sec_seq}  we describe the near optimal sequential and simultaneous sales, and in Appendix \ref{sec:public_nondisc} we show how to achieve a near optimal sale using non-discriminatory prices. %The missing proofs of  this section are deferred to Appendix \ref{app:full-proofs}
%
%Unless stated otherwise, in the spirit of assuming the most optimistic benchmark and the most pessimistic pricing, our benchmark is a seller that can post an exponential size price menu and our pricing scheme sets a single take-it-or-leave-ot price $p_i$ per agent.

\begin{comment}

	\subsection{Revenue Equivalence and Implications}
	\input{equivalence}
\end{comment}
\subsection{Equilibrium and revenue equivalence}\label{sec:public_equi}

	In this section we focus on the case where all agents derive their entire value from a good if purchased by any agent. We first characterize the equilibrium condition for a simultaneous sale. Given an equilibrium $\TT=(T_1,\ldots,T_n)$, the expected value agent $i$ derives from other agents is $\E_{S\sim\mu_\TT^{-i}}\left[x_i(S)\right]=1*\Pr\left[\mbox{some agent }j\neq i\mbox{ buys}\right]=1-\Pr\left[\mbox{no agent }j\neq i\mbox{ buys}\right]=1-\prod_{j\neq i} F_j(T_j).$
	Plugging this expression into Eq. $(\ref{eq:eq_condition})$ yields the following equilibrium condition:
	\begin{eqnarray}
		T_i=\frac{p_i}{\prod_{j\neq i} F_j(T_j)} \quad \mbox{for all $i$}. \label{eq:global_sim_eq}
	\end{eqnarray}
	For a given price vector $\p$ and an equilibrium $\TT\in Eq(\F,\p)$, the expected revenue is % given by
%	and plugging the equilibrium condition into the revenue formulation in Eq. $(\ref{eq:rev_sim})$ gives
	\begin{eqnarray}
	\rsim(\F,\p,\TT)=\sum_{i}p_i\left(1-F_i(T_i)\right)\stackrel{(\ref{eq:global_sim_eq})}{=}\sum_{i}T_i\cdot\Big(\prod_{j\neq  i}F_j(T_j)\Big)\cdot\left(1-F_i(T_i)\right).\label{eq:global_rev_sim}
	\end{eqnarray}
%	where the second inequality follows Eq. \ref{eq:global_sim_eq}.
	
	We turn to describe the equilibrium in the sequential sales model.
	In this case, whenever an agent buys an item, no subsequent agent will ever buy an item.
	Therefore, we can assume without loss of generality that the seller sets a single price per agent.
	Let $\p = (p_1, \ldots, p_n)$ denote the vector of offered prices.
	
	We now show how to compute the unique subgame perfect equilibrium of the game.
	When the last agent (agent $n$) is offered a price, her best strategy is to buy if her value exceeds the price; i.e., $T_n=p_n$. When agent $i=n-1, \ldots, 1$ is offered a price, she faces the following tradeoff: if she buys, her utility is $v_i-p_i$. If she does not buy, her utility is $v_i\left(1-\prod_{j>i}\Pr\left[j\mbox{ does not buy}\right]\right)=v_i\left(1-\prod_{j>i}F_j(T_j)\right)$. Consequently, the unique equilibrium $\TT$ is given by\footnote{Unlike the simultaneous model, an equilibrium exists for non atomless distributions, whenever tie-breaking is done consistently by agents. That is, agents always take the same action when their value is equal to their threshold.}\footnote{For $n$, we let $\prod_{j>n}F_j(T_j)=1$.}
	\begin{eqnarray}
	T_i=\frac{p_i}{\prod_{j>i}F_j(T_j)}\quad \forall i\in [n].\label{eq:eq_condition_seq}
	\end{eqnarray}
	 Given a product distribution $\F$, a price vector $\p$, and an arrival order $\sigma$, let $\tr(\sigma,\p)$ be the function that returns the unique equilibrium.
	 %If the seller sets the prices for the agents knowing the order in which the agents arrive, we say that the seller is \emph{order-aware}; otherwise, we say that the seller is \emph{order-oblivious}.
	 %
	 Since every price vector $\p$ defines a unique strategy vector $\TT$, the expected revenue from agent $i$ is also uniquely defined, and can be calculated by
\begin{eqnarray*}\prod_{j<i}\Pr\left[j \mbox{ does not buy}\right]\cdot p_i \cdot \left(1-F_i(T_i)\right) & \stackrel{(\ref{eq:eq_condition_seq})}{=} & \Big({\prod_{j<i}{F_j(T_j)}}\Big)\cdot T_i\cdot\Big(\prod_{j>i}F_j(T_j)\Big) \cdot\left(1-F_i(T_i)\right) \\
			& = &  T_i\cdot \Big(\prod_{j\neq i}F_j(T_j)\Big)\cdot (1-F_i(T_i)).
		\end{eqnarray*}
	%where the first equality follows from (\ref{eq:eq_condition_seq}).
	Therefore, the expected revenue from all agents can be written as
	\begin{eqnarray}
	\rseq(\F,\sigma,\p,\TT=\tr(\sigma,\p))=\sum_{i}T_{i}\cdot\Big(\prod_{j\neq i}{F_j(T_j)}\Big)\cdot\left(1-F_{i}(T_{i})\right).\label{eq:rev_seq}
	\end{eqnarray}
	Given an arrival order $\sigma$, let $\rbestseq(\F,\sigma)=\max_{\p}\rseq(\F,\sigma,\p,\TT=\tr(\sigma,\p))$ denote the highest revenue a seller can obtain. %We also denote by $\rbwseq(\F)=\max_{\p}\min_{\sigma}\rseq(\F,\sigma,\p,\TT=\tr(\sigma,\p))$ the highest revenue an \emph{order-oblivious} seller can obtain.
	We note that given a threshold vector $\TT$ and an arrival order $\sigma$, there is also a unique price vector that produces this threshold vector $\TT$, which can be calculated by (\ref{eq:eq_condition_seq}), thus $\tr(\sigma,\cdot)$ is a bijection. This is cast in the following observation.
\BOBS\label{obs:seq_single_price}
	Fix an arrival order. An equilibrium strategy vector $\TT$ is uniquely determined by a price vector $\p$, and a price vector $\p$ is uniquely determined by a strategy vector $\TT$.
	\EOBS
	
	The following theorem establishes revenue equivalence in simultaneous and sequential sales.
	
	%	a globally public good,	which means that the underlying network is a clique ($G=K_n$). We first show that a seller who knows in advance the arrival order of the agents in the sequential model yields the same revenue as a seller who can lead the agents to the best equilibrium in the simultaneous model.
	\BT\label{thm:rev_equivalence}
	For every product distribution $\mathcal{F}$ and for every order of arrival $\sigma$ in the sequential model, we have that $\rbestseq(\F,\sigma)=\rbestsim(\F)$.
	\ET
	\BPF
		%Since $\rbestseq(\F,\sigma)$ is defined for an order aware agent,
		%Recall that we assume by renaming that the agents arrive according to  their index.
		By renaming, assume that the agents arrive according to their index. We prove both directions.
		\\\\$\rbestseq(\F)\leq \rbestsim(\F):$ \\
		Let $\p^{seq}$ be a vector in $\argmax_{\p}\rseq(\F,\p,\TT=\tr(\p))$ and let $\tr(\p^{seq})=\TT^{seq}=\left( T^{seq}_1,\ldots,T^{seq}_n\right)$ be the strategies used by the agents for the price vector. From (\ref{eq:rev_seq}) we get that $\rbestseq(\F)=\sum_{i}T^{seq}_i\cdot\prod_{j\neq i}{F_j(T^{seq}_j)}\cdot\left(1-F_i(T^{seq}_i)\right)$.
		According to Observation \ref{obs:sim_single_price}, there exists a single price vector which yields this strategy vector in the simultaneous sales model. We denote this price vector by $\p^{sim}$. From (\ref{eq:global_rev_sim}), we get that $\rsim(\F,\p^{sim},\TT^{seq})=\sum_{i}T^{seq}_i\cdot\prod_{j\neq i}{F_j(T^{seq}_j)}\cdot\left(1-F_i(T^{seq}_i)\right)$. Since $\rsim(\F,\p^{sim},\TT^{seq})\leq \rbestsim(\F)$, the inequality follows.
		\\\\$\rbestsim(\F)\leq \rbestseq(\F):$ \\
		Let $\p^{sim}$ be a price vector in $\argmax_{\p}\rmax(\F,\p)$ and let $\TT^{sim}=\left( T^{sim}_1,\ldots,T^{sim}_n\right)$ be the strategy vector that maximizes the revenue for this price vector, i.e., $\TT^{sim}\in \argmax_{\TT}\rsim(\F,\p^{sim},\TT)$. From (\ref{eq:global_rev_sim}) we have $\rbestsim(\F)= \rsim(\F,\p^{sim},\TT^{sim})=\sum_{i}T^{sim}_i\cdot\prod_{j\neq i}{F_j(T^{sim}_j)}\cdot\left(1-F_i(T^{sim}_i)\right)$. According to Observation (\ref{obs:seq_single_price}), given an order of arrival of the agents, the strategy vector $\TT^{sim}$ is uniquely determined by a price vector $\p^{seq}=\tr^{-1}(\TT^{sim})$ in the sequential sales model. From (\ref{eq:rev_seq}) we have $\rseq(\F,\p^{seq},\TT^{sim})=\sum_{i}T^{sim}_i\cdot\prod_{j\neq i}{F_j(T^{sim}_j)}\cdot\left(1-F_i(T^{sim}_i)\right)$. Since $\rseq(\F,\p^{seq},\TT^{sim})\leq \rbestseq(\F)$, the inequality follows.
	\EPF
	\newline

	%The proofs of Theorem \ref{thm:rev_equivalence} is deferred to Appendix \ref{app:missing_clique}.
	It immediately follows that the optimal revenue is independent of the arrival order.
    \BC
	For every two arrival orders $\sigma, \sigma'$, $\rbestseq(\F,\sigma)=\rbestseq(\F,\sigma')$.
	\EC	
In the sequel, we use $\rbestseq(\F)$ to denote the optimal revenue in the sequential model.

We next draw a connection between selling public goods and selling a single private good. 
This connection is later used in proving approximation results for mechanisms for the sale of public goods. 
Let $Myer(\F)$ denotes the optimal revenue a seller can obtain by selling a single private good to a set of agents drawn from $\F$ (\textit{i.e.}, the revenue obtained by Myerson's optimal auction). Using similar arguments to ones used in \cite{feldman2013pricing}, we have the following:
\BL\label{thm:revbound}
		For every product distribution $\F$, $\rbestseq(\F)\leq Myer(\F)$ \big(and therefore, $\rbestsim(\F)\leq Myer(\F)$ by Theorem \ref{thm:rev_equivalence}\big).
	\EL
	\BPF
	Let $\p^{seq}$ be a vector in $\argmax_{\p}\rseq(\F,\p,\TT=\tr(\p))$ and let $\tr(\p^{seq})=\TT^{seq}=\left( T^{seq}_1,\ldots,T^{seq}_n\right)$ be the equilibrium induced by the price vector  $\p^{seq}$. We have that:
	\begin{eqnarray*}
		\rbestseq(\F) &\stackrel{(\ref{eq:rev_seq})}{=} & \sum_{i}T^{seq}_i\cdot\prod_{j\neq i}{F_j(T^{seq}_j)}\cdot\left(1-F_i(T^{seq}_i)\right)\\
		& \leq & \sum_{i}T^{seq}_i\cdot\prod_{j< i}{F_j(T^{seq}_j)}\cdot\left(1-F_i(T^{seq}_i)\right)\leq Myer(\F),
	\end{eqnarray*}
	where the last inequality is due to the fact that the LHS expression is the expected revenue out of a sequential posted prices mechanism where agent $i$ is offered a price $T^{seq}_i$; this is bounded by the expected revenue of the optimal mechanism for selling a single item.
	\EPF
	
	\subsection{Near optimal simultaneous sale}\label{sec:public_sim}
	In our construction, we use the \textit{ex-ante relaxation} (EAR) \cite{Alaei11,DBLP:journals/corr/AlaeiHNPY15} for selling a private good.
	The EAR relaxes the feasibility constraint, so that instead of selling at most one item \textit{ex post}, this constraint holds only in expectation.
	Since the feasible region increases, the revenue of an optimal mechanism for this case can only be higher than Myerson's optimal mechanism. Combined with Lemma~\ref{thm:revbound}, it suffices to provide a pricing scheme for our setting that approximates the revenue of the EAR.
	As it turns out, when agents' values are drawn from regular distributions, the optimal mechanism for the ex-ante setting
	is a posted price mechanism. These prices can be computed in polynomial time by a convex programming formulation \cite{hartlineMDnA}.
	
	We use these prices to determine prices for the sale of public goods.
	To do so, we partition the agents into {\em valuable} and {\em non-valuable} agents, based on their contribution to the revenue of the EAR. All the revenue obtained in our pricing scheme comes from the valuable agents.
	Their prices are set so that if there exists a valuable agent that buys with low probability, then the equilibrium condition guarantees that the other agents buy with a sufficiently high probability.
	
	%We price the non-valuable agents at $\infty$, and the valuable agents at some scaled-down price relative to their prices in the EAR.
	%We distinguish between two cases: (1) If in the induced equilibrium of our game, all valuable agents buy with at least their corresponding probability in the EAR, then our pricing scheme loses only a constant factor relative to the EAR.
	%(2) Otherwise, there exists a valuable agent who buys in our game with a lower probability than in the EAR.
	%In this case, we utilize the equilibrium condition to infer that the item is bought (by anyone) with a sufficiently high probability.
	%We next state the theorem followed by its full proof.
	
	%	\BT \label{thm:exante-posted prices}
	%	For any regular product distribution $\F$ one can compute in polynomial time prices $\hat{\p}$ that maximize the revenue of an \textit{ex-ante} mechanism.
	%	\ET	
	%	 We show how to transform these  prices into prices that approximate the optimal revenue for the sale of public goods in the simultaneous model.
		\BT
		For social goods with full externalities and for any regular product distribution $\F$, %$\rbwsim(\F)=\Theta(1) \cdot\rbestsim(\F)$. Moreover,
		there exists a poly-time algorithm that computes prices $\p$ for which $\rmin(\F,\p)\geq \nicefrac{\rbestsim(\F)}{5.83}$.\label{thm:exante_reduction}
		\ET
			\BPF
			%Let $\hat{p}$ be the prices that maximizes the revenue of the ex-ante relaxation of the product distribution $\F$ in the private model. We compute $\p$ as follows.
			Let $\hat{p}=(\hat{p}_1,\ldots,\hat{p}_n)$ be the posted prices that maximize the revenue in the EAR, and let $\RR=\sum_i \hat{p}_i (1-F_i(\hat{p}_i))$ be the optimal revenue of the EAR. As mentioned above, $\RR\geq Myer(\F)$.
			Let $c_1,c_2 >1$ be two parameters, to be determined later. We partition the agents into two groups as follows.
			Let $B=\{i\in [n]: \hat{p}_i\geq \RR/c_1\}$ and $S=[n]\setminus B$. For every agent $i$ we set $$p_i=\begin{cases}
			\hat{p}_i/c_2\quad & i\in B\\
			\infty \quad & i\in S
			\end{cases}.$$
			The revenue from the agents in $S$ in the optimal EAR mechanism is bounded by $\sum_{i\in S}\hat{p}_i\cdot\pr\left[i\mbox{ buys}\right]\leq\frac{\RR}{c_1} \sum_{i\in S}(1-F_i(\hat{p}_i))\leq \frac{\RR}{c_1},$ where the last inequality stems from the fact that the EAR sells at most $1$ item in expectation. Therefore, the revenue extracted from agents in $B$ in the EAR is \begin{eqnarray}
			\sum_{i\in B} \hat{p}_i\cdot (1-F_i(\hat{p}_i))\geq \RR-\frac{\RR}{c_1}=\left(1-1/c_1\right)\RR. \label{eq:bigrev_exante}
			\end{eqnarray}
			Let $\TT$ be an equilibrium induced by the price vector $\p=(p_1, \ldots, p_n)$. We consider two cases:
			
			\noindent {\bf Case 1:} $T_i\leq \hat{p}_i$ for every $i\in B$. In this case,
			\begin{eqnarray*}
				\rsim(\F,\p,\TT)& =& \sum_{i}p_i\cdot\left(1-F_i(T_i)\right) = \sum_{i\in B}p_i\cdot\left(1-F_i(T_i)\right)\\
				& \geq & \sum_{i\in B}\frac{\hat{p}_i}{c_2}\cdot\left(1-F_i(\hat{p}_i)\right)
				\stackrel{(\ref{eq:bigrev_exante})}{\geq}
				\left(\frac{1-1/c_1}{c_2}\right)\RR,
			\end{eqnarray*}
			where the first inequality follows from case 1 and the monotonicity of $F_i$. % and the inequality follows from Equation $(\ref{eq:bigrev_exante})$.
			
			\noindent {\bf Case 2:}
			There exists $i\in B$ such that $T_i> \hat{p}_i$. For such an agent $i$,
			\begin{comment}
			$\frac{\hat{p}_i/c_2}{\prod_{j\neq i}F_j(T_j)}=\frac{p_i}{\prod_{j\neq i}F_j(T_j)} \stackrel{(\ref{eq:global_sim_eq})}{=} T_i>\hat{p}_i$. This gives us:
			\begin{eqnarray}
			\prod_{j}F_j(T_j) \leq \prod_{j\neq i}F_j(T_j)\leq \frac{1}{c_2}.\label{eq:buyprob_exante}
			\end{eqnarray}
			\end{comment}
			\begin{eqnarray}
			\frac{\hat{p}_i/c_2}{\prod_{j\neq i}F_j(T_j)}=\frac{p_i}{\prod_{j\neq i}F_j(T_j)} \stackrel{(\ref{eq:global_sim_eq})}{=} T_i>\hat{p}_i
			& \Rightarrow &
			\prod_{j}F_j(T_j) \leq \prod_{j\neq i}F_j(T_j)\leq \frac{1}{c_2}.\label{eq:buyprob_exante}
			\end{eqnarray}
			Let $p_{\min}=\min_{i}p_i$. The expected revenue in this case is at least
			\begin{eqnarray*}
				p_{\min}\cdot\pr[\mbox{at least one agent buys}] %& = & p_{\min}\cdot(1-\pr[\mbox{no agent buys}])\\
				& \geq & \frac{\RR}{c_1 c_2}\Big(1-\prod_{j}F_j(T_j)\Big)
				\stackrel{(\ref{eq:buyprob_exante})}{\geq}  \left(\frac{1-1/c_2}{c_1 c_2}\right)\RR,
			\end{eqnarray*}
			where the first inequality follows from the fact that all prices are at least $\frac{\RR}{c_1 c_2}$. %, and the second inequality follows from Equation $(\ref{eq:buyprob_exante})$.	

			Therefore, we get an approximation factor of $\min\left\{\left(\frac{1-1/c_1}{c_2}\right),\left(\frac{1-1/c_2}{c_1 c_2}\right)\right\}$. Setting $c_1=\sqrt{2}$ and $c_2=1+\frac{1}{\sqrt{2}}$ optimizes the approximation ratio and gives revenue of at least a $\frac{1}{3+2\sqrt{2}}$ fraction of $\RR$. Since $\RR\geq Myer(\F)\geq \rbestsim(\F)$ (by Lemma \ref{thm:revbound}), we get that $\rmin(\F,\p)\geq \frac{\rbestsim(\F)}{3+2\sqrt{2}}\approx\frac{\rbestsim(\F)}{5.83}$. %, as desired.
			\EPF
			
			\vspace{0.1in}
			
			\noindent {\bf Remark:} An approximation ratio of $8$ is given in \cite{feldman2013pricing} for the special case of i.i.d. distributions. The last theorem improves the approximation ratio to $5.83$ even for the more general case of non-identical distributions. Moreover, in Appendix \ref{app:lowerbound} we give a non-discriminatory pricing that gives $4$ approximation for the case of identical distributions. We also show that no pricing scheme can give better approximation than $4/e$, even for identical distributions.
	
%	\begin{comment}
%	Using a similar arguments to the ones presented above, we can also find prices which approximate the optimal revenue when agents arrive simultaneously, even in the case where the agents' valuations are drawn from an \textit{irregular} product distribution $\F$. The proof of the next theorem is deferred to Appendix \ref{app:missing_clique}.
%	\BT
%	For a globally public good and for any (possibly irregular) product distribution $\F$, one can compute a price vector $\p$ for which $\rmin(\F,\p)\geq \rbestsim(\F)\cdot \Theta(1)$.\label{thm:irregular_reduction}
%	\ET
%	\end{comment}
	\subsection{Near optimal sequential sale} \label{sec_seq}
	In this section, we devise a posted price mechanism that approximates the optimal revenue, given a regular product distribution $\F$ and the order of arrival of the agents $\sigma$. The celebrated \textit{prophet inequalities} \cite{krengel1978semiamarts} implies that when selling a single private good, one can find posted prices that achieve an expected revenue of at least half of the optimal mechanism. The posted prices are computed as follows. Let $\phi_i(v)=v-\frac{1-F_i(v)}{f_i(v)}$ be the virtual value function of agent $i$\footnote{See Appendix \ref{app:singleparam} for an overview of the optimal mechanism design theory we refer to.}. Let $t$ be such that $\Pr_{\mathbf{v}\sim \mathcal{F}}\left[\max_i\phi_i(v_i)>t\right]=1/2$, and set $\piprice_i=\phi_i^{-1}(t)$ to be the posted price for agent $i$. By the monotonicity of virtual values in regular distributions, we have that the item is sold with probability of exactly $1/2$, meaning that $\Pr\left[\mbox{item is not sold}\right]=\prod_i F_i(\piprice_i)=1/2$. Assuming that agents arrive according to their index, the result of \cite{krengel1978semiamarts} implies that the expected revenue from posting prices $\piprices=(\piprice_1,\ldots,\piprice_n)$ is
	\begin{eqnarray}
		\sum_i \Pr[i\mbox{ is offered}]\cdot\piprice_i\cdot(1-F_i(\piprice_i)) & = & \sum_i \piprice_i\cdot(1-F_i(\piprice_i))\cdot \prod_{j<i} F_j(\piprice_j) \nonumber\\
		& \geq & Myer(\F)/2,\label{eq:prophineq}
	\end{eqnarray}
	where the equality follows since agent $i$ is offered if and only if no agent arriving before agent $i$ buys the good. We now show that by setting prices that induce thresholds equal to the prophet inequalities prices, the seller obtains a $4$ approximation to the optimal revenue.
	 %a strategy $T_i=\piprices_i$ for every agent $i$		
	%Let $\piprices$ the prices used by the prophet inequality mechanism, as described in Section \ref{section:posted_prices}.
	%We now show how to compute a price vector that approximates the maximal revenue a seller can get in the sequential model.
	%We have the following:
	\BT \label{thm:canonical-rev}
	For goods with full externalities, any regular product distribution $\F$
	%For a globally public good, any regular product distributions $\F$
	 and any known arrival order of the agents $\sigma$, there exists a poly-time algorithm for computing prices $\p$ for which
	$\rseq(\F,\sigma,\p) \geq \rbestseq(\F)/4.$%,\TT=\tr(\sigma,\p))\geq \rbestseq(\F)/4$.
	\ET
	\BPF
		As before, let $t$ be such that $\Pr_{\mathbf{v}\sim \mathcal{F}}\left[\max_i\phi_i(v_i)>t\right]=1/2$. We want to set thresholds such that for every $i$, $T_i=\phi^{-1}_i(t)=\piprice_i$. Therefore, agent $i$ buys if and only if no agent $j< i$ purchased the item, and the virtual value of agent $i$ is greater than $t$. According to Observation \ref{obs:seq_single_price}, knowing the order of arrival, one can find a price vector that supports these thresholds using Eq. (\ref{eq:eq_condition_seq}) (and this price vector is unique). Let $\p=\tr^{-1}(\TT)$ be that vector. Notice that setting the thresholds as described above yields that $\prod_{i\in[n]} F_i(T_i)=1/2$. As shown in Eq. (\ref{eq:rev_seq}), the expected revenue obtained by setting the price vector $\p$ (which supports these thresholds) is:
		\begin{eqnarray}
		 \rseq(\F,\p,\TT)& = &\sum\limits_{i=1}^n T_i\cdot (1-F_i(T_i)) \cdot \prod\limits_{j\neq i} F_j(T_j) \nonumber\\& = &\sum\limits_{i=1}^n T_i\cdot (1-F_i(T_i)) \cdot \prod\limits_{j< i} F_j(T_j)\cdot \prod\limits_{j> i} F_j(T_j).\label{eq:revterms}
		\end{eqnarray}
		Since $\phi_i(T_i)=t$ for every $i$ and for a virtual value $t$ that satisfies the conditions for the prophet inequality, we have that
	\begin{eqnarray}
		\sum\limits_{i=1}^n T_i\cdot (1-F_i(T_i)) \cdot \prod\limits_{j< i} F_j(T_j) = \sum\limits_{i=1}^n \piprice_i\cdot (1-F_i(\piprice_i)) \cdot \prod\limits_{j< i} F_j(\piprice_j) \stackrel{(\ref{eq:prophineq})}{\geq} Myer(\mathcal{F})/2. \label{eq:pi_rev}
	\end{eqnarray}
%	where the inequality is due to Eq. $(\ref{eq:prophineq})$.
	We conclude:
	\begin{eqnarray*}
		\rseq(\F,\p,\TT) & \stackrel{(\ref{eq:revterms})}{=} & \sum\limits_{i=1}^n T_i\cdot (1-F_i(T_i)) \cdot \prod\limits_{j< i} F_j(T_j)\cdot \prod\limits_{j> i} F_j(T_j)\\
		&\geq & \sum\limits_{i=1}^n T_i\cdot (1-F_i(T_i)) \cdot \prod\limits_{j< i} F_j(T_j) /2 \\
		& \stackrel{(\ref{eq:pi_rev})}{\geq} & Myer(\F)/4
		\ \ \ \stackrel{(\mbox{{\tiny Lemma \ref{thm:revbound}}})}{\geq} \ \ \  \rbestseq(\F)/4,
	\end{eqnarray*}
	where the first inequality follows from $\prod_{i\in[n]} F_i(T_i)=1/2$. %, the second inequality follows from $(\ref{eq:pi_rev})$, and the last inequality follows from Theorem \ref{thm:revbound}.
	\EPF
	
	\subsection{Non-discriminatory prices}\label{sec:public_nondisc}
	
	We now show how to approximate the optimal expected revenue of a seller by offering a single, non-discriminatory price, both for the worst case equilibrium in the simultaneous model and for any order of arrival of agents in the sequential model. In the context of selling a single private good, an \textbf{anonymous posted price mechanism} posts a single price $p$, and the first agent to arrive who is willing to pay $p$ obtains the item at price $p$. This is a sequential posted price mechanism that posts the same price $p$ for every agent.
	%In Appendix \ref{app:missing_clique},
	We show that a seller that offers an item to every agent at price $p/2$ obtains at least $1/4$ of the revenue obtained in the case of an anonymous sale of a single private item at price $p$. This holds both in the simultaneous and the sequential cases.
	
	\BT
	Let $\RR=p\cdot (1-\prod_i F_i(p))$ be the revenue of an anonymous price mechanism for the sale of a single private good which posts a price $p$. For a price vector $\p'=\left( \frac{p}{2},\ldots,\frac{p}{2}\right)$, we have:
	\begin{enumerate}[(i)]
		\item $\rmin(\F,\p')\geq \RR/4.$
		\item For every $\sigma$, $\rseq(\F,\sigma,\p',\tr(\sigma, \p'))\geq \RR/4.$
	\end{enumerate} \label{thm:homog_prices}
	\ET\label{thm:quarter_anonymous}
	\BPF
	First, given an equilibrium strategy vector $\TT$, we have that the revenue in the sequential model, for every arrival order is exactly $\frac{p}{2}\cdot (1-\prod_i F_i(T_i))$, while in the simultaneous model, the revenue is at least $\frac{p}{2}\cdot (1-\prod_i F_i(T_i))$, since more than one agent might purchase an item. Similarly to the proof of Theorem \ref{thm:exante_reduction}, we consider two cases:
	\begin{enumerate}
		\item $T_i\leq p$ for every agent $i$: In this case, since the CDFs are monotonically non-decreasing, we have that $\frac{p}{2}\cdot (1-\prod_i F_i(T_i))\geq \frac{p}{2}\cdot (1-\prod_i F_i(p))=\RR/2$.
		\item There exists an agent $i$ such that $T_i> p$: In the simultaneous case, by the equilibrium condition $(\ref{eq:eq_condition})$, we have that $T_i=\frac{p/2}{\prod_{j\neq i} F_j(T_j)}>p$. This implies that $\prod_j F_j(T_j)< \frac{1}{2}$. In the sequential case, for any arrival order $\sigma$, by the equilibrium condition $(\ref{eq:eq_condition_seq})$, we have that $T_i=\frac{p/2}{\prod\limits_{j:\sigma(j)>\sigma(i)} F_j(T_j)}>p$, which implies that $\prod_j F_j(T_j)\leq \prod\limits_{j:\sigma(j)>\sigma(i)} F_j(T_j)<\frac{1}{2}$. In both cases, we get that $\frac{p}{2}\cdot (1-\prod_i F_i(T_i))\geq p/4$. Since $\RR= p\cdot (1-\prod_i F_i(p))\leq p$, we get the  desired result.
	\end{enumerate}
	\EPF

	Applying the $e$-approximation anonymous posted price mechanism presented in \cite{DBLP:journals/corr/AlaeiHNPY15}, we get: % that it is possible to approximate the optimal revenue using non-discriminatory prices, in \textit{both} the sequential and simultaneous models.
	
	\BC \label{cor:nondisc}
	For goods with full externalities and for any product distribution $\F$, there exists a poly-time algorithm for computing a non-discriminatory price $p$ such that:
	\begin{enumerate} [(i)]
		\item $\rmin(\F,\p=\langle p,\ldots, p\rangle)\geq \rbestsim(\F)/4e$; and
		\item For  every order $\sigma$, $\rseq(\F,\sigma,\p,\tr(\sigma, \p))\geq\rbestseq(\F)/4e$.
	\end{enumerate}
	\EC

	\section{Pricing goods with status-based externalities}\label{sec:semi-public}
	Recall that in this setting, every agent is associated with a discount factor $w_i\in [0,1]$, Let $\w=(w_1,\ldots, w_n)$. We next present near-optimal simultaneous and sequential sales.
\begin{comment}The utility of an agent $i$ under price $p_i$ and purchasing set $S \subseteq [n]$ is:
	\begin{eqnarray}
			u_i(S,p_i)=\begin{cases} v_i-p_i \quad & \mbox{if } i\in S\\
			w_i\cdot v_i \quad & \mbox{if }i\notin S\mbox{ and } S\neq \emptyset\\
			0\quad & \mbox{otherwise}\end{cases}.
		\end{eqnarray}
		\end{comment}
		%Given the discount factors of the agents $\w=(w_1,\ldots, w_n)$, our goal is to find posted prices with an expected revenue that approximates the optimal revenue a seller can get by posted prices. We show how to do this for \textit{both} the sequential and simultaneous arrival models.
		%Note that the case of $w_i=0$ and $w_i=1$ coincide with the private goods model and the perfectly public goods model, respectively.
		\subsection{Near optimal sequential sale} \label{sec:fading_seq}%\label{SEC:FADING_SEQ} %
				%We consider a setting where agents arrive sequentially in a predetermined order. We assume by renaming that agents arrive according to their index.
				We devise a non-adaptive pricing scheme (\textit{i.e.}, where an agent's price does not depend on the previous purchases) that approximates the revenue of the optimal adaptive pricing  scheme.\footnote{Which sets a price for the current agent depending on the set of agents that purchased the good prior her arrival.} In our scheme, every agent is assigned with a single price.
		
		%In this setting, the seller posts a price vector $\p$, where $p_i$ is the price offered to agent $i$.
		%Our benchmark is the optimal revenue obtained by a seller that can set a different price for every agent $i$ for every set of agents buying an item before agent $i$ arrives; \textit{i.e.}, the seller posts a set of functions $\hat{\p}=(\hat{p}_1,\ldots,\hat{p}_n)$ in which $\hat{p}_i:2^{[i-1]}\rightarrow\mathbb{R}^{\geq 0}$ sets a different price for any subset of agents who bought the good prior to the arrival of agent $i$.
		%This might impose an exponential strategy space for the agents; nevertheless, our simple pricing approximates this benchmark and reduces the strategy space of the agents.
		
		%As a first step, we show that we can assume that the seller uses only two prices per agent --- a price for the case no agent bought a good prior to the agent's arrival and a price for the case there exists an agent who bought a good before the agent arrived. Then, by posting a single price per agent, the seller can obtain a constant fraction of the optimal revenue she can obtain using two prices.
		
		%Let $\p^0$ and $\p^{>0}$ be the price vectors posted by the seller who uses two price vectors, where $p_i^0$ (resp., $p_i^{>0}$) is the price offered to agent $i$ when no agent (resp., at least one agent) has purchased a good prior to $i$'s arrival. Let $\p=(\p^0,\p^{>0})$.
		
		Let $\p^0$ and $\p^{>0}$ be the price vectors posted by the seller who uses two price vectors, where $p_i^0$ (resp., $p_i^{>0}$) is the price offered to agent $i$ when no agent (resp., at least one agent) has purchased a good prior to $i$'s arrival. Let $\p=(\p^0,\p^{>0})$.			
		The next theorem, proved in Appendix \ref{app:semi_seq}, establishes that it is without loss of generality to restrict attention to two price vectors.
		
		%Given a posted price mechanism $\hat{\p}$, let $\rev(\hat{\p})$ be the expected revenue obtained by the mechanism. The following theorem, whose proof is deferred to Appendix \ref{app:semi_seq}, shows that we can assume that the optimal sale posts two price vectors.
		\BT
		Let $\hat{\p}=(\hat{p}_1,\ldots,\hat{p}_n)$ be a posted-price mechanism where $\hat{p}_i:2^{[i-1]}\rightarrow\mathbb{R}^{\geq 0}$. For every  $\hat{\p}$ there exists a simple mechanism, described by two vectors $\p=(\p^0,\p^{>0})$, for which $\rev(\p)\geq \rev(\hat{\p})$. \label{thm:simple_prices_semi}
		\ET
		In contrast to the full externalities settings, agent $i$ may have two different thresholds in the equilibrium --- one for the case where no agent bought a good before she arrives, denoted by $T_i^0$, and one for the case where at least one agent buys the good, denoted by $T^{>0}_i$.
		For every agent $i$, if some agent bought the good before she arrived, she faces the following trade-off --- if she buys the good, her utility is $v_i-p^{>0}_i$; otherwise, her utility is $w_i\cdot v_i$. Therefore, the threshold satisfies the following equation:
		\begin{eqnarray}
		T^{>0}_i -  p_i^{>0} = w_i \cdot T^{>0}_i \Rightarrow p_i^{>0}= (1-w_i)\cdot T^{>0}_i. \label{eq:bought_threshold}
		\end{eqnarray}
		If no agent bought the good before before agent $i$ arrived, then\footnote{For the case of $i=n$, the RHS product is naturally defined to be $1$, and therefore $T_n^0=p_n^0$.}
		\begin{eqnarray}
		& &T_i^0 - p_i^0  =  w_i \cdot T_i^0 \cdot \pr[\mbox{Agent $j > i$ buys a good}] 
		=  w_i \cdot T_i^0 \cdot \Big(1 - \prod_{j>i}{F_j(T_j^0)}\Big)\nonumber\\
		& &\Rightarrow p_i^0 = (1-w_i) \cdot T_i^0 + w_i \cdot T_i^0 \cdot \prod_{j>i}{F_j(T^0_j)}.\label{eqn:fading_threshold}
		\end{eqnarray}
		
For every agent $i$ and pricing $\p=(\p^0,\p^{>0})$, let $q_i^0=q_{i}^0(\p)$ (resp., $q_i^{>0}=q_i^{>0}(\p)$) denote the probability that no agent (resp., at least one agent) has bought a good before agent $i$ arrived. %(and therefore, agent $i$ is offered price $p_i^0$)
%, and $q_i^{>0}=q_i^{>0}(\p)$  as the probability that at least one previous agent has bought a good. Therefore, the revenue can be written as %(and in this case, agent $i$ is offered price $p_i^{>0}$). For ease of notation, we drop $\p$, and simply write $q_{i}^0$ and $q_i^{>0}$. The revenue of the auction can be expressed by
The revenue can now be written as		\begin{eqnarray}
			\rev(\p) & = & \sum_i \left(q_i^0\cdot p_i^0\cdot (1-F_i(T_i^0))+q_i^{>0}\cdot p_i^{>0}\cdot (1-F_i(T_i^{>0}))\right) \nonumber\\
			& \stackrel{(\ref{eqn:fading_threshold})}{=} & \sum_i q_i^0\cdot \Big((1-w_i) \cdot T_i^0 + w_i \cdot T_i^0 \cdot \prod_{j>i}{F_j(T^0_j)}\Big)\cdot (1-F_i(T_i^0))+\sum_i q_i^{>0}\cdot p_i^{>0}\cdot (1-F_i(T_i^{>0})) \nonumber\\
			& \stackrel{(\ref{eq:bought_threshold})}{=} & \sum_i q_i^0\cdot (1-w_i) \cdot T_i^0 \cdot (1-F_i(T_i^0))+
			\sum_i q_i^0\cdot w_i \cdot T_i^0 \cdot \Big(\prod_{j>i}{F_j(T^0_j)}\Big) \cdot (1-F_i(T_i^0))\nonumber\\
			 & & + \sum_i q_i^{>0}\cdot T_i^{>0}\cdot (1-w_i)\cdot (1-F_i(T_i^{>0})). \nonumber
	\end{eqnarray}		
By removing some occurrences of factors smaller than 1 ($q_i^0, q_i^{>0}, w_i$) in the last expression, we get
\begin{eqnarray}
\rev(\p) & \leq & \sum_i  (1-w_i) \cdot T_i^0 \cdot (1-F_i(T_i^0))+
			\sum_i  \Big(\prod_{j<i} F_j(T_j^0)\Big)\cdot  T_i^0 \cdot \Big(\prod_{j>i}{F_j(T^0_j)}\Big) \cdot (1-F_i(T_i^0))\nonumber\\
			 & & + \sum_i T_i^{>0}\cdot (1-w_i)\cdot (1-F_i(T_i^{>0})) \nonumber\\
			& = & \sum_i  (1-w_i) \cdot T_i^0 \cdot (1-F_i(T_i^0))+ \sum_i (1-w_i)\cdot T_i^{>0}\cdot  (1-F_i(T_i^{>0})) \nonumber\\
			& & + \sum_i T_i^0 \cdot \Big(\prod_{j\neq i}{F_j(T^0_j)}\Big) \cdot (1-F_i(T_i^0)). \label{eq:revbound_semi_seq}
		\end{eqnarray}
		%where the inequality holds since we removed factors smaller than 1.
		
		Given a thresholds vector $\TT=(T_1,T_2,\ldots, T_n)$, we define $\rev_1(\TT,\w)  =  \sum_i  (1-w_i) \cdot T_i \cdot (1-F_i(T_i))$ and $\rev_2(\TT)  =  \sum_i T_i \cdot \left(\prod_{j\neq i}{F_j(T_j)}\right) \cdot (1-F_i(T_i))$.
		It follows from Eq. $(\ref{eq:revbound_semi_seq})$ that
		\begin{eqnarray}
		\max_{\p}\rev(\p)\leq 2\max_{\TT}\rev_1(\TT,\w) + \max_{\TT}\rev_2(\TT).\label{eq:revbound_fading_seq}
		\end{eqnarray}
		
		That is, the RHS sum in Eq. $(\ref{eq:revbound_fading_seq})$ is an upper bound on the optimal revenue that can be obtained. $\rev_1(\TT,\w)$ can be viewed as the private component of the revenue, which becomes more significant as $w_i$'s get smaller, while $\rev_2(\TT)$ can be viewed as the public component, which becomes more significant as $w_i$'s grow. Notice that $\max_{\TT}\rev_2(\TT)$ is exactly $\rbestseq(\F)$, where $\rbestseq(\F)$ is the optimal posted prices revenue in a sequential sale in the full externalities model, as defined in Section \ref{sec:clique}.
		
		We now show that it is possible to find prices that approximate $\max_{\TT}\rev_1(\TT,\w)$ and prices that approximate $\max_{\TT}\rev_2(\TT)$. In fact, we show a stronger result, namely that for each of the terms in the sum, there exists a single price vector $\p=\p^0=\p^{>0}$ that approximates it. This is shown in Lemmas \ref{lm:rev_priv} and \ref{lm:rev_pub} whose proofs are deferred to Appendix \ref{app:semi_seq}.
		\BL
			There exists a poly-time algorithm for computing prices $\p$ such that $\rev(\p)\geq \max_{\TT}\rev_1(\TT,\w)$.\label{lm:rev_priv}
		\EL

			Recall that $\rseq(\F,\p')$ is the expected revenue from posting price vector $\p'$ to agents that arrive sequentially in the full externalities model. %, and that $\rbestseq(\F)$ is the optimal attainable revenue for that model.}
			The following allows us to reduce the problem of finding ``good" prices in the status-based externalities model to finding ``good" prices in the full externalities model.

		\BL
			Given prices $\p'$, there exist poly-time computable prices $\p$ such that $\rev(\p)\geq \rseq(\F,\p')$.\label{lm:rev_pub}
			%Given that there exist prices $\p'$ there exist poly-time computable prices $\p$ such that $\rseq(\F,\p')\leq \rev(\p)$.
		\EL
		
		We now present the main result of this section:
		
		\BT 
			Given a $c$-approximation pricing for sequential sales in the full externalities model, there exists a poly-time computable pricing that guarantees a $(c+2)$-approximation for the optimal sequential sales in the model of status-based externalities.\label{thm:semi_seq_main}
		\ET
		\BPF
			Since $\max_{\TT}\rev_2(\TT)=\rbestseq(\F)$, if one can find prices that $c$-approximate the optimal prices in the full externalities model, by Lemma \ref{lm:rev_pub}, one can compute prices that $c$-approximate $\max_{\TT}\rev_2(\TT)$ in the status-based externalities model.
			
			Let $\p_1$ and $\p_2$ be the sets of prices for which $\rev(\p_1)\geq \max_{\TT}\rev_1(\TT,\w)$ and $c\cdot\rev(\p_2)\geq \max_{\TT}\rev_2(\TT)$, respectively.
These prices can be computed in poly time by Lemmas \ref{lm:rev_priv} and \ref{lm:rev_pub}. 
We have that
			\begin{eqnarray*}
				\max_{\p}\rev(\p) & \stackrel{(\ref{eq:revbound_fading_seq})}{\leq} & 2\max_{\TT}\rev_1(\TT,\w) + \max_{\TT}\rev_2(\TT)\\
				&\leq & 2\cdot \rev(\p_1) + c\cdot \rev(\p_2)\\
				&\leq & (c+2)\cdot \max\{\rev(\p_1),\rev(\p_2)\}.
			\end{eqnarray*}
		\EPF

	The following corollary directly follows from Theorems \ref{thm:semi_seq_main} and \ref{thm:canonical-rev}.
	\BC\label{cor:status_seq_approx}
		For goods that exhibit status-based externalities, there exists a poly-time algorithm for computing prices that give a $6$-approximation to the optimal pricing scheme.
	\EC

		\begin{comment}
		\BT
		For goods that exhibit social status-based externalities, There exists a poly-time algorithm for computing prices $\p=(p_1,\ldots,p_n)$ that approximate the optimal posted prices sale to within a constant factor in the sequential model. \label{thm:semi_seq_main}
		\ET
		\BPF
			Follows directly from Eq. $(\ref{eq:revbound_fading_seq})$, Lemma \ref{lm:rev_priv} and Lemma \ref{lm:rev_pub}.
		\EPF
		\end{comment}
		We note that unlike in the full externalities scenario, price discrimination is needed. This is since the revenue obtained by using non-discriminatory prices can be smaller by a factor of $\Theta(\log n)$ than the optimal revenue, as shown in Appendix \ref{app:inapprox_homog}.

		\subsection{Near optimal simultaneous sale}\label{sec:fading_sim}
				%We consider a setting where agents are offered prices simultaneously. The seller posts a price vectors $\p$ where $p_i$ is the price offered to agent $i$. 
		
		Given an equilibrium threshold vector $\TT=(T_1,\ldots,T_n)$, the expected value agent $i$ derives from other agents is $\E_{S\sim\mu_\TT^{-i}}\left[x_i(S)\right]=w_i\cdot \Pr\left[\mbox{some agent $j\neq i$ buys}\right]=w_i\left(1-\prod_{j\neq i}F_j(T_j)\right).$  
		Plugging this expression into Eq. (\ref{eq:eq_condition}) we get that in equilibrium, for every $i$
		%Let us characterize the set of equilibria in this model. Given the threshold vector $\TT_{-i}$ of all agents but $i$, if agent $i$ buys a good, her utility is $v_i-p_i$; otherwise, if agent $i$ does not buy the good, her expected utility can be expressed by $v_i\cdot w_i (1-\prod_{j\neq i}F_j(T_j))$. We get that in equilibrium, for every $i$:
		\begin{eqnarray}
			p_i=(1-w_i)\cdot T_i + w_i\cdot T_i\cdot \prod_{j\neq i} F_j(T_j). \label{eq:equilibrium_fading_sim}
		\end{eqnarray}

	Our main result of this section is:
	\BT
		For goods that exhibit status-based externalities, there exists a poly-time algorithm for computing prices for a simultaneous sale that approximate the optimal posted prices sale to within a constant factor at any equilibrium. \label{thm:semi_sim_main} %\label{THM:SEMI_SIM_MAIN}%\label{thm:semi_sim_main}
	\ET
	
	In order to prove Theorem \ref{thm:semi_sim_main}, we show that the revenue, $\rev(\p,\TT)$, that is obtained by prices $\p$ at equilibrium $\TT$ is upper bounded by $\rev_1(\TT,\w) + \rev_2(\TT)$, where $\rev_1$ and $\rev_2$ are as defined in Section \ref{sec:fading_seq}. We then show how to devise prices that obtain a constant factor approximation to $\max_{\hat{\TT}}\rev_1(\hat{\TT},\w)$ at any equilibrium, and prices that obtain a constant factor approximation to $\max_{\hat{\TT}}\rev_2(\hat{\TT})$ at any equilibrium. Ergo, we devise prices that approximate the better of the two, which in turn, give a constant factor approximation to the optimal attainable revenue.
	The full details of the proof of Theorem \ref{thm:semi_sim_main} are given in Appendix \ref{app:semi_sim}.

	\section{Pricing goods with availability-based externalities} \label{sec:semi-gradual}%\label{SEC:SEMI-GRADUAL}
	%\subsection{Sequential Model}
Recall that in this scenario, the fraction agent $i$ gets when a set $S\not  \owns i$ purchases the good is $w(\abs{S})$ for a non-decreasing function $w:[n]\rightarrow [0,1]$.
\begin{comment}The utility of an agent $i$ under price $p_i$ and purchasing set $S \subseteq [n]$ is:
	\begin{eqnarray}
	u_i(S,p_i)=\begin{cases} v_i-p_i \quad & \mbox{if } i\in S\\
	w(\abs{S})\cdot v_i \quad & \mbox{if }i\notin S\\
	0\quad & \mbox{otherwise}\end{cases}.
	\end{eqnarray}
	\end{comment}
 For ease of notation, we use $w_k=w(k)$ and normalize $w_n=1$ and $w_0=0$. For each agent $i$, the seller sets a vector of prices $p_i=(p_i^0,\ldots,p_i^{i-1})$, where $p_i^j$ is the price that the seller sets for agent $i$ if $j$ agents bought a good prior to her arrival; let $\p$ be the price matrix $(p_1,\ldots, p_n)$.
Our benchmark is the optimal revenue obtained by any posted price matrix $\p$.

Our main result of this section, is the following: %whose proof is deferred to Appendix \ref{app:grad_analysis}, is:
\BT
	For goods that exhibit availability-based externalities, there exists a poly-time algorithm for computing a price matrix $\p$ for sequential sale that approximates the optimal sale using a price matrix to within an $O(\log n)$ factor. \label{thm:grad_main}
\ET

We first describe the equilibrium obtained by the agents when the seller posts a  matrix $\p$.
%Let $\p=(p_1,\ldots,p_n)$ be a price matrix posted by the seller, where $p_i=(p_i^0,\ldots, p_i^{i-1})$ is the price vector posted to agent $i$.
For every agent $i$, we define a set of thresholds $T_i^0,\ldots, T_i^{i-1}$, where $T_i^j(\p)$ is the minimal value for which agent $i$ buys the good \textit{conditioned} on $j$ agents buying the good prior to her arrival given the price matrix $\p$. We use the following notations:
\begin{enumerate}
	\item $q_{i,j}(\p)$: the probability that $j$ agents buy a good prior to agent $i$'s arrival, given prices $\p$.
	\item $r_{i,j}^k(\p)$: the probability that \textbf{exactly} $k$ agents buy a good given prices $\p$ and given that $j$ agents bought a good prior to agent $i$'s arrival. For ease of notations, we define $r_{n+1,j}^j=1$ and $r_{n+1,j}^k=0 $ for every $j \neq k$.
	\item $r_{i,j}^{< k}(\p) = \sum_{\ell=j}^{k-1}r_{i,j}^{\ell}$: the probability that \textbf{strictly less} than $k$ agents buy a good overall given prices $\p$ and given that $j$ agents bought a good prior to $i$'s arrival. 
\end{enumerate}
When clear from the context, we omit $\p$ and simply write $T_i^j, q_{i,j}, r_{i,j}^k$ and $r_{i,j}^{< k}$. Given these notations, we now describe the equilibrium condition for agent $i$. If agent $i$ buys a good when $j$ agents buy a good prior to her arrival, her utility from buying a good at price $p_i^j$ is $v_i-p_i^j$. On the other hand, if the agent does not buy a good, her expected utility is $v_i\cdot \sum_{\ell\geq j} w_\ell\cdot r_{i+1,j}^\ell$. Therefore, the threshold from which it is better for agent $i$ to buy an item when $j$ agents bought an item prior to agent $i$ satisfies:
\begin{eqnarray*}
	T_i^j-p_i^j=T_i^j\cdot \sum_{\ell\geq j} w_\ell\cdot r_{i+1,j}^\ell \Rightarrow p_i^j= T_i^j(1-\sum_{\ell\geq j} w_\ell\cdot r_{i+1,j}^\ell )
\end{eqnarray*}
Clearly, if $j$ agents bought the good prior to agent $i$'s arrival, then for every $\ell<j$, $r_{i+1,j}^\ell =0$. Therefore, we can rewrite the above condition as follows: 
\begin{eqnarray}
p_i^j & = & T_i^j(1-\sum_{\ell\geq j} w_\ell\cdot r_{i+1,j}^\ell )\ = \ T_i^j(w_n\cdot\sum_{\ell\geq j} r_{i+1,j}^{\ell}-\sum_{\ell\geq j} w_{\ell}\cdot r_{i+1,j}^\ell )\nonumber\\
& = & T_i^j\cdot\sum_{\ell\geq j} r_{i+1,j}^{\ell}(w_n- w_\ell) \ =\  T_i^j\cdot\sum_{\ell\geq j} r_{i+1,j}^\ell \cdot\sum_{k>\ell}(w_{k}-w_{k-1})\nonumber\\
& = & T_i^j\cdot\sum_{k>j}(w_{k}-w_{k-1})\cdot\sum_{\ell=0}^{k-1} r_{i+1,j}^\ell \ =\ T_i^j\cdot\sum_{k>j}(w_{k}-w_{k-1})\cdot r_{i+1,j}^{< k},\label{eq:gradual_seq_eq}
\end{eqnarray}
where the second equality stems from $w_n=1$ and $\sum_{\ell\geq j} r_{i+1,j}^\ell =1$.

Notice that for $j\geq i$, $q_{i,j}=0$.
We therefore can express the revenue from setting prices $\p$ as:
\begin{eqnarray}
\rev(\p)& = &\sum_{i,j} q_{i,j} \cdot p_i^j\cdot(1-F_i(T_i^j))\nonumber\\
& = & \sum_{i,j} q_{i,j}\cdot T_i^j\cdot \sum_{k>j}(w_{k}-w_{k-1})\cdot r_{i+1,j}^{<k}\cdot(1-F_i(T_i^j))\nonumber\\
& = & \sum_{k=1}^{n}(w_{k}-w_{k-1})\cdot \sum_{i}\sum_{j< k} q_{i,j}\cdot T_i^j\cdot (1-F_i(T_i^j))\cdot r_{i+1,j}^{< k}.\label{eq:grad_seq_eq_rev}
\end{eqnarray}

For a given vector of thresholds $\TT$ and a given $k$, let $$\revpubseqmult(\TT)=\sum_{i}\sum_{j<k} q_{i,j}\cdot T_i^j\cdot (1-F_i(T_i^j))\cdot r_{i+1,j}^{< k}.$$
In order to bound the revenue in Eq. $(\ref{eq:grad_seq_eq_rev})$, we draw a connection between $\revpubseqmult(\TT)$ and the sale of $k$ identical items without externalities.
As we define in Appendix \ref{app:singleparam}, let $\RR(\F,\feas_k,\sigma,\p)$ be the revenue of a posted price mechanism for agents drawn from a product distribution $\F$, and offered prices $\p$ sequentially, according to order $\sigma$, until $k$ items are sold in a setting with no externalities ($\feas_k$ is the $k$-uniform matroid feasibility constraint). Since we assume that agents arrive according to their index, we omit $\sigma$ from now on. We denote by $\RR^*(\F,\feas_k)$ the revenue of the optimal posted price mechanism for selling $k$ private items sequentially in the same order of arrival as in our social goods setting.  
We establish the following lemma:
\BL\label{lm:revpubmultbound}
For every threshold vector $\TT$ and every number of items $k$, 
\begin{eqnarray*}
	\revpubseqmult(\TT)\leq \RR^*(\F,\feas_k). 
\end{eqnarray*}
\EL
\BPF
Let us define the following posted prices sale for selling $k$ items privately. We define the price matrix $\hat{\p}$, where 
$${\hat{p}}_i^j=\begin{cases} T_i^j\quad & j< k\\ \infty\quad & \mbox{otherwise}\end{cases}.$$ 
When agent $i$ arrives, the seller posts a price of ${\hat{p}}_i^j$ if $j$ goods where purchased prior to agent $i$'s arrival.

For $j< k$, let  $\hat{q}_{i,j}$ be the probability that $j$ items are sold prior agent $i$'s arrival. We show by induction on $i$ that for every $i$ and $j$ $\hat{q}_{i,j}=q_{i,j}$. For the first agent to arrive (\textit{i.e.}, $i=1$), clearly $q_{i,0}=\hat{q}_{i,0}=1$ (and obviously, $q_{i,j}=\hat{q}_{i,j}=0$ for every other $j$). Let us assume that the claim holds for agent $i$ (and for every $j$). It is clear that for $j< k$, $$q_{i+1,j}=q_{i,j-1}\cdot (1-F_{i}(T_{i}^{j-1}))+q_{i,j}\cdot F_i(T_i^j)={\hat{q}}_{i,j-1}\cdot (1-F_{i}({\hat{p}}_{i}^{j-1}))+{\hat{q}}_{i,j}\cdot F_i({\hat{p}}_i^j)={\hat{q}}_{i+1,j}.$$

Therefore, the revenue from selling $k$ items in a private sale using prices $\hat{\p}$ can be expressed by:
\begin{eqnarray*}
	\sum_{i}\sum_{j< k} {\hat{q}}_{i,j} \cdot {\hat{p}}_i^j \cdot (1-F_i({\hat{p}}_i^j)) & = &  \sum_{i}\sum_{j<k} {q}_{i,j} \cdot {T}_i^j \cdot (1-F_i({T}_i^j))\\
	&\geq & \sum_{i}\sum_{j<k} q_{i,j}\cdot T_i^j\cdot (1-F_i(T_i^j))\cdot r_{i+1,j}^{< k}\\
	& = & \revpubseqmult(\TT).
\end{eqnarray*}
Since selling $k$ items using prices $\hat{\p}$ is a private posted price sale of at most $k$ items, the revenue obtained by this sale is at most that of the optimal posted price sale for selling $k$ items, $\RR^*(\F,\feas_k)$. Therefore, we get the desired result.
\EPF

For $k>n$, we set $w_k=1$. Using Lemma \ref{lm:revpubmultbound}, we can bound Eq. $(\ref{eq:grad_seq_eq_rev})$ and get
\begin{eqnarray*}
	\rev(\p)& \leq& \sum_{k=1}^{n}(w_{k}-w_{k-1})\cdot \RR^*(\F,\feas_k)\\
	&= &\sum_{\ell=0}^{\lfloor\log n\rfloor}\sum_{k=2^\ell}^{2^{\ell+1}-1}(w_{k}-w_{k-1})\cdot \RR^*(\F,\feas_k)\\
	& \leq & \sum_{\ell=0}^{\lfloor\log n\rfloor}\sum_{k=2^\ell}^{2^{\ell+1}-1}(w_{k}-w_{k-1})\cdot \RR^*\left(\F,\feas_{\min(2^{\ell+1}-1,n)}\right)\\
	& = & \sum_{\ell=0}^{\lfloor\log n\rfloor}(w_{2^{\ell+1}-1}-w_{2^{\ell}-1})\cdot \RR^*\left(\F,\feas_{\min(2^{\ell+1}-1,n)}\right),
\end{eqnarray*}
where the second equality stems from $(w_{k}-w_{k-1})\cdot \RR^*(\F,\feas_k)=(1-1)\cdot \RR^*(\F,\feas_k)=0$ for $k>n$. 

In order to show an $O(\log n )$ approximation to the optimal sequential posted price sell, we show how to produce a sale of revenue $\Omega(1)\cdot (w_{2k+1}-w_{k})\cdot \RR^*(\F,\feas_{2k+1})$ for every $k\leq n $. This is given in the two following lemmas, whose proofs are deferred to Appendix \ref{app:grad-proofs}.

\BL
For every $k\in \{1,\ldots,n\}$, there exist prices $\p$ that guarantee an expected revenue of $\Omega(1)\cdot (w_{2k+1}-w_{k})\cdot \RR^*(\F,\feas_{2k+1})$.\label{lem:grad1}
\EL
\BL
There exist prices $\p$ that guarantee an expected revenue of $\Omega(1)\cdot \RR^*(\F,\feas_{1})$.\label{lem:grad2}
\EL

The proof of Theorem \ref{thm:grad_main} follows by combining the upper bound on the optimal revenue in Eq. \eqref{eq:grad_log} and the approximation guarantees from Lemmas \ref{lem:grad1} and \ref{lem:grad2}.

	\section{Hardness results for network-based externalities}\label{sec:obsevations}
		In this section we show that unless $\mathrm{P}=\mathrm{NP}$, for network-based externalities and for every $\epsilon>0$, we cannot output a price vector for which the revenue is an $O(n^{1-\epsilon})$-approximation to the optimal revenue. This is true in the simultaneous model, even when all agents' valuations are drawn from the uniform distribution over $[0,1]$. In the sequential model, this is true even when agents all have a fixed value 1. Recall that in the network-based externalities model, externalities are modeled by a graph, and agent derives her entire value if she or one of her neighbors by the good; that is, $x_i(S)=1$ if and only if $S\cap \{i\cup N(i)\}\neq \emptyset$, where $N(i)$ denotes the neighbors of $i$ in the graph. 
	In order to show our hardness result, we use the following hardness result obtained by Zuckerman:
	\BT \cite{zuckerman2006linear}
		For all  $\epsilon> 0$, it is NP-hard to approximate the size of the maximal independent set in a graph to within a factor of $O(n^{1-\epsilon})$.\label{thm:max-is-nphard}
	\ET 
	We first prove the hardness result for the simultaneous model:
	\BT
		Assume that all agents’ valuations are drawn i.i.d. from the uniform distribution over $[0, 1]$. Given a (undirected) graph $G$ with $n$ nodes, it is NP-hard to compute prices which approximate the optimal revenue when agents arrive simultaneously to within a factor of $O(n^{1-\epsilon})$. \label{thm:hard_sim}
	\ET
	\BPF
		Let $k$ be the size of the maximum independent set in $G=(V=[n],E)$. We reduce the problem of  approximating the size of the maximal independent set in $G$ to the problem of maximizing revenue for agents in $[n]$ who exhibit network-based externalities based on the graph $G$. We first describe the equilibrium condition. Given agents' strategies $\TT$, we have that $\E_{S\sim \mu_{\TT}^{-i}}\left[x_i(S)\right]=1-\prod_{j\in N(i)} F_j(T_j)$. Therefore, Eq. $(\ref{eq:eq_condition})$ becomes $p_i = T_i\cdot\prod_{j\in N(i)} F_j(T_j)$ for every $i$. %, and the expected revenue can be written as 
%		$$\rsim(\F,\p,\TT)=\sum_i T_i\cdot \left(\prod_{j\in N(i)} F_j(T_j)\right)\cdot (1-F_i(T_i)).$$
	
		We show that by computing prices $\p$ for which $\rmin(\F,\p)$ approximates  $\rbwsim(\F)= \max_{\p^*}\rmin(\F,\p)$ to within a factor of $O(n^{1-\epsilon})$, we can approximate the size of the independent set in $G$ to within a factor $O(n^{1-\epsilon})$ as well, which is NP-hard according to Theorem \ref{thm:max-is-nphard}. We note that $\rbwsim(\F)\leq \rbestsim(\F)$, which implies we cannot approximate $\rbestsim(\F)$ as well.
	
		By fixing a maximum size independent set $S$, and offering a price of $1/2$ to each agent $i$ in $S$ and $\infty$ to all agents not in $S$, we get a revenue of $1/4$ from agent $i$ for every agent in $S$. This is true since none of agent $i$'s neighbors purchase a good, and therefore $T_i=p_i=1/2$. Therefore, we get that $\rbwsim(\F)\geq k/4$.
		
		%We show a reduction from the maximum independent set problem. 
		We now show the reduction. We are given an algorithm that takes the graph as input and computes a price vector $p$ satisfying $\rmin(\p)\geq \rbwsim(\F)/\Theta(n^{1-\epsilon})$ for some constant $\epsilon>0$. 
		We compute an equilibrium using the following: 
		\begin{figure}[h]\label{fig:equilibrium}
		\MyFrame{	
			\begin{enumerate}
				\item[] Assume by renaming that $p_1 \leq p_2 \leq \ldots \leq p_n$.
				\item[] $S \gets \emptyset$
				\item[] For $i=1,\ldots,n$:
				\begin{itemize}
					\item[] If $N(i) \cap S =\emptyset$:
					\begin{itemize}
						\item[] $S=S \cup \{i\}$
						\item[] $T_i = p_i$
						\end{itemize}		
						\item[] Else:
						\begin{itemize}
							\item[] $T_i = 1$
							\end{itemize}		
							
							\end{itemize}
							\end{enumerate}
							} %\caption{A process for finding an equilibrium when all agents are distributed uniformly on [0,1].}			
		\end{figure}
		
		We show that for every agent $i$, $T_i$ is the best-response.
		We distinguish between two cases:
		\begin{enumerate}
			\item $i \in S$: it is clear that for every $j \in N(i), F_j(T_j)=1$ and therefore $T_i=\frac{p_i}{\prod_{j \in N(i)}F_j(T_j)}=p_i$.
			\item $i \notin S$: there exists $k<i$ such that $k \in N(i) \cap S$ and therefore $$T_i=\frac{p_i}{\prod_{j \in N(i)}F_j(T_j)} \geq\frac{p_k}{F_k(T_k)} =\frac{p_k}{F_k(p_k)} = 1,$$ which is equivalent to having a threshold $T_i= 1$.   
		\end{enumerate}
		Notice that in the equilibrium computed, the set $S$ is an independent set. This follows since in each step we check whether adding $i$ into $S$ preserve that $S$ is still an independent set. Therefore, $\rsim(\F,\p,\TT) \leq k$. 
		We output $|MAX-IS|=	\rsim(\F,\p,\TT)=\sum_i p_i\cdot(1-F_i(T_i))$. If the algorithm finds prices such that $\rmin(\F,\p)\geq\frac{\rbestsim(\F)}{O(n^{1-\epsilon})}$, we get we computed a number in $\left[\frac{k}{O(n^{1-\epsilon})},k\right]$, which is $\mathrm{NP}$-hard according to Theorem \ref{thm:max-is-nphard}.			
	\EPF
	
The following theorem shows that it is hard for the seller to compute prices that maximize the seller's revenue, even in the case where all agents' valuations are known to be equal to $1$. 
\BT
	Assume that all agents’ have a fixed value of $1$ and that tie-breaking is done-consistently. Given a graph $G$ with $n$ nodes, it is NP-hard to compute prices which approximate the optimal revenue when agents arrive sequentially to within a factor of $O(n^{1-\epsilon})$. \label{thm:hard_seq}
\ET
\BPF
Let $k$ be the size of the maximum independent set in $G$. By fixing a maximum size independent set, and offering a price of $1/2$ to each agent $i$ in the set, we get a revenue of $1/2$ from agent $i$ (and a price $>1$ for all other agents). This is true since none of agent $i$'s neighbors purchase a good. Therefore, we get that $\rbestseq(\F)\geq k/2$. Since we assume tie-breaking is done consistently, we can assume that $v_i\neq p_i$ for every $i$.
We now show how to compute an equilibrium:
\\\\
\MyFrame{	
	\begin{enumerate}
		\item[] $S \gets \emptyset$
		\item[] For $i=n,\ldots,1$:
		\begin{itemize}
			\item[] If $(p_i< v_i)$ and $N(i) \cap S =\emptyset$:
			\begin{itemize}
				\item[] $S=S \cup \{i\}$
			\end{itemize}		
		\end{itemize}
	\end{enumerate}
}
\\\\
We claim that in a subgame perfect equilibrium, only agents in $S$ buy a good. We show it by induction on the number of agents. For $n=1$, this is trivial, since the single agent buys if and only if $p_i<v_i$. Let us assume the claim is true for $n<n'$, and show the claim for $n=n'$. Let us inspect the first agent to arrive. We inspect two cases:
\begin{enumerate}
	\item The first agent does not buy a good (\textit{i.e.}, the first agent is not in $S$): this implies one of the following. 
	\begin{itemize}
		\item $v_1<p_1$. In this case the agent has a negative utility of purchasing the good and he would not buy. 
		\item  A neighboring agent purchases a good and the first agent has a utility of $v_1$, which is greater than $v_1-p_1$, the utility of the agent if she were to purchase a good. After the first agent does not buy a good, this is equivalent to a network without the first agent, and by the inductive assumption, the rest of the players are in a subgame perfect equilibrium. This is, one of the neighbors of the first agent buys a good. Therefore, the first agent gets a higher utility in the subgame perfect equilibrium induced by the agent not buying a good.
	\end{itemize} 
	We get that in this case, the first agent does not buy in an equilibrium. Since this is equivalent to a network without the first agent, by the inductive assumption, the rest of the players are in a subgame perfect equilibrium when set $S$ is the purchasing set.
	
	\item The first agent buys a good (the first agent is in $S$). This implies that $v_1>p_1$ and that non of the agent's neighbors is in $S$ ($N(1)\cap S=\emptyset$). Therefore, if the first agent buys a good, her utility is $v_1-p_1>0$.
	If the first agent were not to buy a good, this would have been equivalent to a network without the first agent. By the inductive assumption, in an equilibrium, the set $S-\{1\}$ would  have bought a good. Therefore, non of agent 1's neighbors would have bought a good, and her utility would have been $0$. This implies that in equilibrium, the first agent buys.
	
	Whenever the first agent buys a good, her neighbors do not buy a good, and this is equivalent to a network without the first agent and the agents in $N(1)$. It is not hard to verify that in this case, the computation above returns the set $S-\{1\}$ as a purchasing set. This follows since the non-purchasing agents have no effect on the agents that arrive prior to them or after them in the computation. That is, the set $S$ buys overall in the equilibrium where the first agent buys. 
\end{enumerate}
We get that in both cases, the subgame perfect equilibrium is for agents in set $S$ to purchase a good, which completes the inductive proof.

Notice that in the equilibrium computed above, the set $S$ of purchasing agents is an independent set, and therefore, $\sum_{i\in S} p_i\leq \abs{S}\leq k$. Our reduction from maximum independent set computes $S$ and returns $\abs{MAX-IS} = \sum_{i\in S} p_i$. We conclude that if the algorithm finds prices such that $\rseq(\F, \p)\geq \frac{\rbestseq(\F)}{O(n^{1-\epsilon})}$, we can compute a number in $\left[\frac{k}{O(n^{1-\epsilon})},k\right]$, which is $\mathrm{NP}$-hard according to Theorem \ref{thm:max-is-nphard}.
\EPF

	% Renew this after \maketitle if the default list of authors is too long for headers
	%\renewcommand{\shortauthors}{W.\ Vickrey et.\ al.}
	
	%\section{Missing proofs of Section \ref{sec:fading_seq}}\label{app:semi_seq}

	% Bibliography
	%\bibliographystyle{ACM-Reference-Format}
	
	\bibliography{public}

\begin{thebibliography}{10}

\bibitem{ahmadipouranari2013equilibrium}
Nima AhmadiPourAnari, Shayan Ehsani, Mohammad Ghodsi, Nima Haghpanah, Nicole
  Immorlica, Hamid Mahini, and Vahab Mirrokni.
\newblock Equilibrium pricing with positive externalities.
\newblock {\em Theoretical Computer Science}, 476:1--15, 2013.

\bibitem{akhlaghpour2010optimal}
Hessameddin Akhlaghpour, Mohammad Ghodsi, Nima Haghpanah, Vahab~S Mirrokni,
  Hamid Mahini, and Afshin Nikzad.
\newblock Optimal iterative pricing over social networks.
\newblock In {\em International Workshop on Internet and Network Economics},
  pages 415--423. Springer, 2010.

\bibitem{Alaei11}
Saeed Alaei.
\newblock Bayesian combinatorial auctions: Expanding single buyer mechanisms to
  many buyers.
\newblock In {\em {IEEE} 52nd Annual Symposium on Foundations of Computer
  Science, {FOCS} 2011, Palm Springs, CA, USA, October 22-25, 2011}, pages
  512--521, 2011.
\newblock URL: \url{http://dx.doi.org/10.1109/FOCS.2011.90}, \href
  {http://dx.doi.org/10.1109/FOCS.2011.90} {\path{doi:10.1109/FOCS.2011.90}}.

\bibitem{DBLP:journals/corr/AlaeiHNPY15}
Saeed Alaei, Jason~D. Hartline, Rad Niazadeh, Emmanouil Pountourakis, and Yang
  Yuan.
\newblock Optimal auctions vs. anonymous pricing.
\newblock In {\em {IEEE} 56th Annual Symposium on Foundations of Computer
  Science, {FOCS} 2015, Berkeley, CA, USA, 17-20 October, 2015}, pages
  1446--1463, 2015.
\newblock URL: \url{http://dx.doi.org/10.1109/FOCS.2015.92}, \href
  {http://dx.doi.org/10.1109/FOCS.2015.92} {\path{doi:10.1109/FOCS.2015.92}}.

\bibitem{calo2013digital}
Ryan Calo.
\newblock Digital market manipulation.
\newblock {\em Geo. Wash. L. Rev.}, 82:995, 2013.

\bibitem{candogan2010optimal}
Ozan Candogan, Kostas Bimpikis, and Asuman Ozdaglar.
\newblock Optimal pricing in the presence of local network effects.
\newblock In {\em International Workshop on Internet and Network Economics},
  pages 118--132. Springer, 2010.

\bibitem{chawla2007algorithmic}
Shuchi Chawla, Jason~D Hartline, and Robert Kleinberg.
\newblock Algorithmic pricing via virtual valuations.
\newblock In {\em Proceedings of the 8th ACM conference on Electronic
  commerce}, pages 243--251. ACM, 2007.

\bibitem{chawla2010multi}
Shuchi Chawla, Jason~D Hartline, David~L Malec, and Balasubramanian Sivan.
\newblock Multi-parameter mechanism design and sequential posted pricing.
\newblock In {\em Proceedings of the forty-second ACM symposium on Theory of
  computing}, pages 311--320. ACM, 2010.

\bibitem{feldman2015combinatorial}
Michal Feldman, Nick Gravin, and Brendan Lucier.
\newblock Combinatorial auctions via posted prices.
\newblock In {\em Proceedings of the Twenty-Sixth Annual ACM-SIAM Symposium on
  Discrete Algorithms}, pages 123--135. SIAM, 2015.

\bibitem{feldman2013pricing}
Michal Feldman, David Kempe, Brendan Lucier, and Renato Paes~Leme.
\newblock Pricing public goods for private sale.
\newblock In {\em Proceedings of the fourteenth ACM conference on Electronic
  commerce}, pages 417--434. ACM, 2013.

\bibitem{haghpanah2013optimal}
Nima Haghpanah, Nicole Immorlica, Vahab Mirrokni, and Kamesh Munagala.
\newblock Optimal auctions with positive network externalities.
\newblock {\em ACM Transactions on Economics and Computation}, 1(2):13, 2013.

\bibitem{hannak2014measuring}
Aniko Hannak, Gary Soeller, David Lazer, Alan Mislove, and Christo Wilson.
\newblock Measuring price discrimination and steering on e-commerce web sites.
\newblock In {\em Proceedings of the 2014 conference on internet measurement
  conference}, pages 305--318. ACM, 2014.

\bibitem{hartline2008optimal}
Jason Hartline, Vahab Mirrokni, and Mukund Sundararajan.
\newblock Optimal marketing strategies over social networks.
\newblock In {\em Proceedings of the 17th international conference on World
  Wide Web}, pages 189--198. ACM, 2008.

\bibitem{hartlineMDnA}
Jason~D. Hartline.
\newblock {\em Mechanism Design and Approximation}.
\newblock 2016.

\bibitem{HR09}
Jason~D. Hartline and Tim Roughgarden.
\newblock Simple versus optimal mechanisms.
\newblock In {\em Proceedings 10th {ACM} Conference on Electronic Commerce
  (EC-2009), Stanford, California, USA, July 6--10, 2009}, pages 225--234,
  2009.
\newblock URL: \url{http://doi.acm.org/10.1145/1566374.1566407}, \href
  {http://dx.doi.org/10.1145/1566374.1566407}
  {\path{doi:10.1145/1566374.1566407}}.

\bibitem{kleinberg2012matroid}
Robert Kleinberg and Seth~Matthew Weinberg.
\newblock Matroid prophet inequalities.
\newblock In {\em Proceedings of the forty-fourth annual ACM symposium on
  Theory of computing}, pages 123--136. ACM, 2012.

\bibitem{krengel1978semiamarts}
Ulrich Krengel and Louis Sucheston.
\newblock On semiamarts, amarts and processes with finite value.
\newblock {\em Advances in Prob}, 4:197--266, 1978.

\bibitem{mas1995microeconomic}
Andreu Mas-Colell, Michael~Dennis Whinston, Jerry~R Green, et~al.
\newblock {\em Microeconomic theory}, volume~1.
\newblock Oxford university press New York, 1995.

\bibitem{myerson1981optimal}
Roger~B Myerson.
\newblock Optimal auction design.
\newblock {\em Mathematics of operations research}, 6(1):58--73, 1981.

\bibitem{samuelson1954pure}
Paul~A Samuelson.
\newblock The pure theory of public expenditure.
\newblock {\em The review of economics and statistics}, pages 387--389, 1954.

\bibitem{zuckerman2006linear}
David Zuckerman.
\newblock Linear degree extractors and the inapproximability of max clique and
  chromatic number.
\newblock In {\em Proceedings of the thirty-eighth annual ACM symposium on
  Theory of computing}, pages 681--690. ACM, 2006.

\end{thebibliography}
	
	\newpage
	\appendix

\section{Brief overview on single parameter environments without externalities} \label{app:singleparam}% and posted price mechanisms}
%\subsection{Single-parameter environments without externalities}\label{section:environments}
\paragraph{\textbf{Selling a private good}.} We draw interesting connections between the posted price mechanism for selling social goods and the nearly optimal mechanism for selling a single good in a setting with no externalities. We denote this as ``selling a private good" or as the ``private model." Next, we give a brief overview of some of the main concepts in this field.
\BD[Virtual values and Regularity]
	Let $F$ be the cumulative distribution function of an atomless distribution, and let $f$ be its corresponding density function. The virtual value function associated with $F$ is  defined as $\phi(v)=v-\frac{1-F(v)}{f(v)}$. The distribution $F$ is regular if its corresponding $\phi$ is non-decreasing.
\ED
In a seminal paper of \cite{myerson1981optimal}, Myerson showed that the mechanism that obtains optimal revenue from selling a private good is a virtual value maximizer; \textit{i.e.}, it sells the good to the agent with the highest virtual value as long as her virtual value is non-negative.
Note that given a product distribution from which agent valuations are drawn, the optimal mechanism induces for every agent a probability for getting served, and the sum of probabilities is no greater than $1$.

\paragraph{\textbf{Ex ante relaxation}.} To achieve good upper bounds in the private model, a useful method is to consider the \textit{ex-ante relaxation}\cite{Alaei11,DBLP:journals/corr/AlaeiHNPY15}. In the private model, if a good is given to one agent, it cannot be assigned to another one. In the \textit{ex-ante relaxation}, a good can be assigned to more than one agent, as long as the expected number of agents assigned is at most one. Clearly, the revenue of such a mechanism is an upper bound of the optimal mechanism that satisfies this condition \textit{ex post} (\textit{e.g.}, Myerson's optimal mechanism). For regular distributions, the optimal \textit{ex-ante} mechanism turns out to be a simple posted price mechanism and can be efficiently computed using a convex programming formulation.

\paragraph{\textbf{Monopoly price}.} Given a single agent whose valuation is drawn from a regular distribution $F$, the \textit{Monopoly Price} is the posted price $p^*$ that maximizes the expected revenue extracted from the agent, $p^*=\argmax_{p}p\cdot (1-F(p))$. Let $\phi$ be the virtual value function for distribution $F$. The monopoly price satisfies $\phi(p^*)=0$.

\paragraph{\textbf{Anonymous price mechanism}.} Such a mechanism is defined in the context of selling a single private good. An \textbf{anonymous posted price mechanism} posts a single price $p$, and the first agent to arrive who is willing to pay $p$ obtains the item at price $p$. This is a sequential posted price mechanism that posts the same price $p$ for every agent. In \cite{DBLP:journals/corr/AlaeiHNPY15}, the authors showed the following:
\BT\cite{DBLP:journals/corr/AlaeiHNPY15}\label{thm:anonymous_e}
For any regular product distribution $\F$, there exists an anonymous price $p$ such that $p\cdot (1-\prod_i F_i(p))\geq Myer(\F)/e$. Furthermore, this price can be computed in polynomial time.\footnote{This polynomial time computation was not explicitly shown in \cite{DBLP:journals/corr/AlaeiHNPY15}. One of the prices offered by the optimal \textit{ex-ante} mechanism is an anonymous price that achieves an $e$-approximation, as clarified in a personal communication with Emmanouil Pountourakis.}
\ET

\paragraph{\textbf{Multi-item settings}.} Myerson's optimal mechanism naturally extends to a setting where multiple \textit{identical} private goods are for sale, and each agent wants a single good (unit-demand settings); simply maximize the sum of virtual values of the agents served and charge them according to Myerson's payment identity \cite{myerson1981optimal}. Chawla et al. \cite{chawla2010multi} showed how to design a posted price mechanism that approximates the revenue of the optimal Myerson mechanism. 

\paragraph{\textbf{Sequential posted price mechanisms}.} Gaining popularity, posted price mechanisms approximate the revenue obtained by optimal mechanisms for a wide variety of settings using a surprisingly simple auction structure; Simply offer a take it or leave it price for agents in a sequential order under a predetermined feasibility constraint. Chawla et al. \cite{chawla2010multi} distinguished between two types of mechanisms. Sequential posted price mechanism (SPM) in which the seller can choose the order in which the agents are offered the prices, and Oblivious posted price mechanisms (OPMs) in which the agents' order of arrival is fixed in advanced and known to the seller. As we note above, they showed how to approximate the optimal revenue of Myerson's optimal auction using an OPM when selling $k$ identical items under a $k$-uniform matroid feasibility constraint. 

\paragraph{\textbf{Nearly optimal OPM for multi-item settings.}} One of the settings addressed by \cite{chawla2010multi} is the multi-item setting, where $k$ items are sold using an OPM, and the feasibility constraint is a $k$-uniform matroid (meaning that every set of size at most $k$ can be served), denoted by $\feas_k$. Let $\RR(\F,\feas_k,\sigma, \p)$ denote the revenue of an OPM using pricing $\p$ when agents arrive according to $\sigma$ with valuations drawn from $\F$ under feasibility constraint $\feas_k$. % valuations are drawn from $\F$  arrive according to order $\sigma$ under constraint $\feas_k$. 
\cite{chawla2010multi} showed the following:
\BT\cite{chawla2010multi} \label{thm:k_uniform_rev}
For any regular product distribution $\F$ and for any order $\sigma$, there exists a polytime algorithm for computing prices $\p$ such that $\RR(\F,\feas_k,\sigma, \p)\geq Myer(\F,\feas_k)/2$, where $Myer(\F,\feas_k)$ denotes the revenue of the Myerson optimal mechanism given the product distribution $\F$ and the feasibility constraint $\feas_k$.
\ET

\section{$4/e$ lower bound and $4$ upper bound for simultaneous sale of public goods for i.i.d. agents} \label{app:lowerbound}
We first show that there exists a set of distributions $\F$ for which no (possibly non-discriminatory) pricing achieves more than $e/4$ fraction of the optimal attainable revenue $\rbestsim(\F)$ at the worst equilibrium.

\BT 
There exists a product of identical regular distributions $\F$ such that for every $\p$, $\rmin(\F,\p)\leq \rbestsim(\F)\cdot \nicefrac{e}{4}$.\label{thm:lb}
\ET 
\BPF
We set $\F$ to be the product of identical distributions, where every $F_i$ is uniform over $[0,1]$. Given a price vector $\p=(p_1, \ldots,p_n)$, the following yields an equilibrium. Assume by renaming that $p_1\leq p_2\leq \ldots \leq p_n$, and assume $p_1\leq 1$ (otherwise, no agent buys). By setting $T_1=p_1$ and $\{T_i=p_i/p_1\geq 1\}_{i>1}$, we have that $F_1(T_1)=p_1$ and for every $i> 1$ $F_i(T_i)=1$. Therefore, $T_1\cdot \prod_{i>1}F_i(T_i)=p_1$ and $T_i\cdot \prod_{j\neq i} F_j(T_j)=\frac{p_i}{p_1}\cdot p_1=p_i$ for $i>1$, and the equilibrium condition $(\ref{eq:global_sim_eq})$ holds. In this equilibrium, no agent $i>1$ buys. Thus, agent $1$ cannot rely on others, and buys the good if and only if $v_1>p_1$. In order to maximize the seller's revenue at this equilibrium, we set $p_1=1/2$ (the monopoly price), which leads to  
\begin{eqnarray}
\rmin(\F,\p)\leq 1/4 \label{eq:minub}
\end{eqnarray}
for every $\p$.

Now consider a seller that sets a non-discriminatory price $p=\left(1-\frac{1}{n+1}\right)^n$ and the symmetric equilibrium $T_i= \left(1-\frac{1}{n+1}\right)$ for every $i$. It is easy to verify that the equilibrium condition $(\ref{eq:global_sim_eq})$ holds. We get that the revenue in this equilibrium is 
\begin{eqnarray}
\sum_i p\cdot(1-F_i(T_i))=n\cdot \left(1-\frac{1}{n+1}\right)^n\cdot\frac{1}{n+1}= \left(1-\frac{1}{n+1}\right)^{n+1}\approx 1/e\leq \rbestsim(\F).\label{eq:maxlb}
\end{eqnarray}
Combining $(\ref{eq:minub})$ with $(\ref{eq:maxlb})$ yields the desired result.
\EPF

In the following, we describe a non-discriminatory pricing that gives a $4$-approximation to the optimal revenue for the case where agents' valuations are drawn i.i.d.. This is achieved using a simple adaptation of the technique used in the proof of Theorem \ref{thm:exante_reduction}. We note that this is a factor 2 improvement over the main result in \cite{feldman2013pricing}.

\BT 
For every product of identical regular distributions $\F$  there exists an algorithm for computing a non-discriminatory pricing $\p$, such that  $\rmin(\F,\p)\geq \rbestsim(\F)/4$.\label{thm:iid}
\ET 
\BPF
Our starting point is the prices maximizing the ex-ante relaxation revenue for $\F$. Since the distributions are identical, the prices are identical, and we denote them by $\hat{p}$. Our non-discriminatory pricing sets a price $p_i=\hat{p}/2$ for every agent $i$. The proof follows similar lines to the proof of Theorem \ref{thm:exante_reduction} by noticing that all agents are in set $B$, and by setting $c_2=2$.
\EPF

\section{Missing proofs of Section~\ref{sec:fading_seq}}\label{app:semi_seq}
%\section{Missing proofs of Section \ref{sec:clique}}\label{app:semi_seq}

\textbf{Proof of Theorem \ref{thm:simple_prices_semi}:}
Recall that $\hat{\p}$ is a posted price mechanism offered by the seller. We show how to transform such a mechanism into a simple mechanism which can be defined by only two price vectors $\p^0$ and $\p^{>0}$ which obtains at least as much revenue.
For some agent $i$ and for some set $S\subseteq [i-1]$, let $\hat{q}_i^S$ denote the probability that exactly the set $S$ of agents purchased the good prior to agent $i$'s arrival in the mechanism defined by $\hat{\p}$ (and agent $i$ was offered price $\hat{p}_i(S)$).

We define the exponential strategy space for agent $i$. Given that a non-empty set $S$ have bought the good prior to agent $i$, the agent faces two choices --- if the agent buys a good, her utility is $v_i-\hat{p}_i(S)$; otherwise, her utility is $w_i\cdot v_i$. Rearranging gives us:
\begin{eqnarray}
\hat{T}_i^{S}=\frac{\hat{p}_i(S)}{1-w_i}. \label{eq:subset_eq_fading}
\end{eqnarray}
Let $\hat{T}_j^{\emptyset}$ be the threshold strategy of some agent $j$ arriving after agent $i$ when offered price $\hat{p}_j(\emptyset)$. We now define the threshold strategy $\hat{T}_i^\emptyset$.
If no agent has bought the good, she again faces two choice --- when buying the good, her utility is $v_i-\hat{p}_i(\emptyset)$; otherwise, her utility is 
\begin{eqnarray*}
	w_i\cdot \pr\left[\mbox{Some agent $j>i$ buys a good}\right] & = & w_i\cdot \left(1-\pr\left[\mbox{No agent $j>i$ buys a good}\right]\right)\\
	& = & w_i\cdot \left(1-\prod_{j>i}F_j(\hat{T}_j^{\emptyset})\right).
\end{eqnarray*}
Rearranging give us: 
\begin{eqnarray}
\hat{T}_i^{\emptyset}=\frac{\hat{p}_i(\emptyset)}{1-w_i +w_i\prod_{j>i}F_j(\hat{T}_j^{\emptyset})}. \label{eq:emptyset_eq_fading}
\end{eqnarray}
We show how to set the price vectors $\p^0$ and $\p^{>0}$. Let $p^{M}_i$ denote the monopoly price for agent $i$,  we set $p_i^0=\hat{p}_i^{\emptyset}$ and $p_i^{>0}=(1-w_i)\cdot p^{M}_i$. We denote $\p=(\p^0,\p^{>0})$. Setting these prices imposes the following threshold strategy for agent $i$:
\begin{itemize}
	\item Whenever no agent buys a good prior to agent $i$'s arrival, her threshold strategy is $T_i^0=\hat{T}_i^\emptyset$.
	\item If an agent buys an item prior to agent $i$'s arrival, her threshold strategy is $T_i^{>0}=p^{M}_i$.
\end{itemize} 
Let $q_i^0$ denote the probability that no agent purchased a good prior to agent $i$'s arrival in the mechanism defined by $\p$ (and agent $i$ being offered a price $p_i^0$). Note that: $$q_i^0=\prod_{j<i}F_j(T_j^0)= \prod_{j<i}F_j(\hat{T}_j^\emptyset)=\hat{q}_i^\emptyset.$$ The probability that agent $i$ is offered a price $(1-w_i)\cdot p^M_i$ is $1-q_i^0= 1-\hat{q}_i^\emptyset$.
We conclude:
\begin{eqnarray*}
	\rev(\hat{\p}) & = & \sum_i\sum_{S\subset [i-1]} \hat{q}_i^S\cdot \hat{p}_i(S)\cdot (1-F_i(\hat{T}_i^S))\\
	& = &  \sum_i\hat{q}_i^{\emptyset}\cdot \hat{p}_i^{\emptyset}\cdot (1-F_i(\hat{T}_i^\emptyset)) + \sum_i\sum_{S\neq \emptyset \subset [i-1]} \hat{q}_i^S\cdot \hat{p}_i(S)\cdot (1-F_i(\hat{T}_i^S))\\
	& = &  \sum_i\hat{q}_i^{\emptyset}\cdot \hat{p}_i^{\emptyset}\cdot (1-F_i(\hat{T}_i^\emptyset)) + \sum_i\sum_{S\neq \emptyset \subset [i-1]} \hat{q}_i^S\cdot (1-w_i)\cdot \hat{T}_i^S\cdot (1-F_i(\hat{T}_i^S))\\
	&\leq & \sum_i\hat{q}_i^{\emptyset}\cdot \hat{p}_i^{\emptyset}\cdot (1-F_i(\hat{T}_i^\emptyset))  + \sum_i\sum_{S\neq \emptyset \subset [i-1]} \hat{q}_i^S\cdot (1-w_i)\cdot p^{M}_i\cdot (1-F_i(p^{M}_i))\\
	& = & \sum_i\hat{q}_i^{\emptyset}\cdot \hat{p}_i^{\emptyset}\cdot (1-F_i(\hat{T}_i^\emptyset))  + \sum_i (1-\hat{q}_i^\emptyset)\cdot (1-w_i)\cdot p^{M}_i\cdot (1-F_i(p^{M}_i))\\
	& = & \sum_i q_i^0\cdot p_i^{0}\cdot (1-F_i(T_i^0)) + \sum_i (1-q_i^{0})\cdot  p_i^{>0}\cdot (1-F_i(T_i^{>0})) \\
	& = & \rev(\p),
\end{eqnarray*}
where the in inequality follows from the monopoly price definition.
\qedsymb\\\\

\textbf{Proof of Lemma \ref{lm:rev_priv}:}	
Let $\hat{\p}$ be a vector such that $\hat{p}_i$ is the monopolist price for distribution $F_i$. Notice that $\hat{\p}=\argmax_{\TT}\rev_1(\TT,\w)$ for every $\w$.
We set prices $\p$ such that for every $i$, $$p_i^0=p_i^{>0}=(1-w_i)\hat{p}_i.$$

We first show that both $T_i^{0}$ and $T_i^{>0}$ are both no bigger than $\hat{p}_i$. By Eq. $(\ref{eq:bought_threshold})$, we get that $T_i^{>0}=\hat{p}_i$. By Eq. $(\ref{eqn:fading_threshold})$, we get: 
$$T_i^0=\frac{p_i^0}{(1-w_i)+w_i\cdot\prod_{j>i} F_j(T_j^0)}\leq  \frac{(1-w_i)\cdot \hat{p}_i}{(1-w_i)}=\hat{p}_i.$$

We get that the revenue from setting prices $\p$ is at least:
\begin{eqnarray*}
	\rev(\p) & = & \sum_i\left( q_i^0\cdot p_i^0\cdot (1-F_i(T_i^0))+q_i^{>0}\cdot p_i^{>0}\cdot (1-F_i(T_i^{>0})) \right)\\
	& \geq & \sum_i \left(q_i^0\cdot (1-w_i)\cdot \hat{p}_i\cdot (1-F_i(\hat{p}_i))+q_i^{>0}\cdot (1-w_i)\cdot \hat{p}_i\cdot (1-F_i(\hat{p}_i))\right) \\
	& = & \sum_i (1-w_i)\cdot \hat{p}_i\cdot (1-F_i(\hat{p}_i))\\
	& = & \rev_1(\hat{\p},\w)\\
	& = &\max_{\TT}\rev_1(\TT,\w),
\end{eqnarray*}
as desired.	\qedsymb\\\\
\textbf{Proof of Lemma \ref{lm:rev_pub}:}
%Note that $\rbestseq(\F)=\max_{\TT} \rev_2(\TT)$. 
Given price $\p'$, %for which $\rseq(\F,\p')\geq \rbestseq(\F)/c$, 
let $\hat{\TT}=\tr(\p')$ be a set of thresholds which induced by $\p'$ in the full externalities model. % $c$-approximate $\rbestseq(\F)$ in the full externalities model (which can be computed  according to Theorem \ref{thm:canonical-rev}), and let $\p'=\tr^{-1}(\hat{\TT})$  be the prices \aenote{that induce} $\hat{\TT}$ as an equilibrium in that model. 
We set prices $\p$ as follows --- for every agent $i$ set: 
\begin{eqnarray*}
	p_i^0  = p_i^{>0} =  (1-w_i)\cdot \hat{T}_i + w_i\cdot\hat{T}_i\cdot \prod_{j>i} F_j(\hat{T}_j),
\end{eqnarray*}
Note that by the way we set prices, according to Eq. (\ref{eqn:fading_threshold}), $T_i^0=\hat{T}_i$ for every $i$.
We get that the expected revenue is:
\begin{eqnarray*}
	\rev(\p) & = & \sum_i\left( q_i^0\cdot p_i^0\cdot (1-F_i(T_i^0))+q_i^{>0}\cdot p_i^{>0}\cdot (1-F_i(T_i^{>0})) \right)\\
	& \geq & \sum_i q_i^0\cdot p_i^0\cdot (1-F_i(T_i^0))\\
	& \stackrel{(\ref{eqn:fading_threshold})}{=} & \sum_i q_i^0\cdot \left((1-w_i)\cdot T_i^0 + w_i\cdot T_i^0\cdot \prod_{j>i} F_j(T_j^0)\right)\cdot (1-F_i(T_i^0))\\
	& \geq & \sum_i q_i^0\cdot \left((1-w_i)\cdot T_i^0\cdot \prod_{j>i} F_j(T_j^0) + w_i\cdot T_i^0\cdot \prod_{j>i} F_j(T_j^0)\right)\cdot (1-F_i(T_i^0))\\
	& = & \sum_i\left(\prod_{j<i}F_j(T_j^0)\right)\cdot T_i^0\left(\prod_{j>i} F_j(T_j^0)\right)\cdot (1-F_i(T_i^0))\\
	& = & \sum_i\left(\prod_{j\neq i}F_j(\hat{T}_j)\right)\cdot \hat{T}_i\cdot (1-F_i(\hat{T}_i))\\
	& = & \rseq(\F,\p') %\geq \rbestseq(\F)/c = \max_{\TT}  \rev_2(\TT)/c.
\end{eqnarray*}
Note that in the above, the third equality follows since the probability that agent $i$ will be offered price $p_i^0$, $q_i^0$, is exactly the probability that no agent buys a good before $i$ arrives, $\prod_{j<i}F_j(T_j^0)$.\qedsymb

\section{Proof of Theorem~\ref{thm:semi_sim_main}}\label{app:semi_sim}
	Recall that $\p=(p_1,\ldots, p_n)$ is the price vector posted by the seller, where $p_i$ is the price offered to agent $i$, and let $\TT$ be some equilibrium induced by $\p$ (therefore, $\TT$ satisfies Eq. $(\ref{eq:equilibrium_fading_sim})$).
		The revenue of the sale when $\p$ are the prices and $\TT$ is equilibrium can be expressed by: 
		\begin{eqnarray}
			\rev(\p,\TT) & = & \sum_i  p_i\cdot (1-F_i(T_i))\nonumber\\
			& \stackrel{(\ref{eq:equilibrium_fading_sim})}{=} & \sum_i \left((1-w_i) \cdot T_i + w_i \cdot T_i \cdot \prod_{j\neq i}{F_j(T_j)}\right)\cdot (1-F_i(T_i)) \nonumber\\
			& = & \sum_i (1-w_i) \cdot T_i \cdot (1-F_i(T_i))+
			\sum_i w_i \cdot T_i \cdot \left(\prod_{j\neq i}{F_j(T_j)}\right) \cdot (1-F_i(T_i))\nonumber\\
			& \leq & \sum_i (1-w_i) \cdot T_i \cdot (1-F_i(T_i))+
			\sum_i T_i \cdot \left(\prod_{j\neq i}{F_j(T_j)}\right) \cdot (1-F_i(T_i))\nonumber\\
			& = & \rev_1(\TT,\w) + \rev_2(\TT)\nonumber\\
			& \leq & \max_{\hat{\TT}}\rev_1(\hat{\TT},\w)+ \max_{\hat{\TT}}\rev_2(\hat{\TT}). \label{eq:rev_bound_fade_sim}
		\end{eqnarray}
		
		We now show that it is possible to find prices that approximate $\max_{\hat{\TT}}\rev_1(\hat{\TT},\w)$ and prices that approximate $\max_{\hat{\TT}}\rev_2(\hat{\TT})$. This yields a constant approximation for our model.

		\BL
			One can compute prices $\p$ such that the expected revenue $\rmin(\p)\geq \max_{\TT}\rev_1(\TT,\w)$.\label{lm:rev_priv_sim}
		\EL
		\BPF
			Let $\hat{\p}$ be a vector such that $\hat{p}_i$ is the monopolist price for distribution $F_i$. Notice that $\hat{\p}=\argmax_{\TT}\rev_1(\TT,\w)$ for every $\w$.
			We set prices $\p$ such that for every $i$, $$p_i=(1-w_i)\cdot \hat{p}_i.$$
			
			We first show that in every equilibrium $\TT$,  $T_i\leq \hat{p}_i$. By Eq. $(\ref{eq:equilibrium_fading_sim})$, we get: 
			$$T_i \stackrel{(\ref{eq:equilibrium_fading_sim})}{=}\frac{p_i}{(1-w_i)+w_i\cdot\prod_{j\neq i} F_j(T_j)}\leq  \frac{(1-w_i)\cdot \hat{p}_i}{(1-w_i)}=\hat{p}_i.$$

			We get that for every $\TT$, the expected revenue from setting prices $\p$ is at least:
			\begin{eqnarray*}
				\rev(\p,\TT) & = & \sum_i p_i\cdot (1-F_i(T_i))\\
				& \geq & \sum_i (1-w_i)\cdot \hat{p}_i\cdot (1-F_i(\hat{p}_i))\\
				& = & \rev_1(\hat{\p},\w)\\
				& = &\max_{\TT}\rev_1(\TT,\w),
			\end{eqnarray*}
			as desired.	
		\EPF
		
		\BL
			One can compute prices $\p$ such that the expected revenue $\rev(\p)\geq \Omega(1)\cdot\max_{\TT}\rev_2(\TT)$.\label{lm:rev_pub_sim}
		\EL
		\BPF		
			Let $\hat{p}$ be an anonymous price for which
			\begin{eqnarray}
				\hat{p}\cdot\left(1-\prod_{i} F_i(\hat{p}_i)\right)=  \Omega(1)\cdot Myer(\F)= \Omega(1) \cdot \rev_2(\F).\label{eq:anonymous_fade_sim}
			\end{eqnarray}
			This uniform price exists and can be computed according to \cite{DBLP:journals/corr/AlaeiHNPY15}.
			We post a uniform price vector $\p$ such $p_i=\hat{p}/2$ for every $i$. 
			Let $\RR=\hat{p}\cdot (1-\prod_i F_i(\hat{p}))$ be the revenue of an anonymous price mechanism which posts a price $p$. 
Given an equilibrium strategy vector $\TT$, we have that the expected revenue is at least $\frac{\hat{p}}{2}\cdot (1-\prod_i F_i(T_i))$, since more then one agent might purchase an item. 
\begin{enumerate}
	\item $T_i\leq \hat{p}$ for every agent $i$: In this case, since the CDFs are monotonically non-decreasing, we have that $\frac{\hat{p}}{2}\cdot (1-\prod_i F_i(T_i))\geq \frac{\hat{p}}{2}\cdot (1-\prod_i F_i(\hat{p}))=\RR/2$.
	\item There exists an agent $i$ such that $T_i> \hat{p}$: By the equilibrium condition $(\ref{eq:equilibrium_fading_sim})$, we have that $T_i=\frac{\hat{p}/2}{(1-w_i)+w_i\cdot \prod_{j\neq i} F_j(T_j)}>\hat{p}$. Which yields $(1-w_i)+w_i\cdot \prod_{j\neq i} F_j(T_j)< \frac{1}{2}$. We get:
	\begin{eqnarray*}
		\prod_{j} F_j(T_j) & \leq & \prod_{j\neq i} F_j(T_j)\\
		& =  &(1-w_i)\cdot  \prod_{j\neq i} F_j(T_j) + w_i\cdot \prod_{j\neq i} F_j(T_j)\\
		& \leq & (1-w_i)+ w_i\cdot \prod_{j\neq i} F_j(T_j)\\
		& < & \frac{1}{2}.
	\end{eqnarray*}
	That is, the expected revenue is at least: $$\frac{\hat{p}}{2}\cdot (1-\prod_i F_i(T_i))>\frac{\hat{p}}{4}= \Omega(1)\cdot Myer(\F)= \Omega(1)\cdot \rev_2(\F),$$ as desired.
\end{enumerate}

		\EPF
		
The proof of Theorem \ref{thm:semi_sim_main} follows from Equation $(\ref{eq:rev_bound_fade_sim})$, and Lemmata \ref{lm:rev_priv_sim} and \ref{lm:rev_pub_sim}.\footnote{This proof gives an approximation factor of $4e+1$. We note that we can get a factor of $\approx 6.83$ using the EAR prices instead of the anonymous price in the proof of Lemma~\ref{lm:rev_pub_sim} via a slightly more involved proof.}

\section{Missing proofs of Section~\ref{sec:semi-gradual}}\label{app:grad-proofs}

\noindent\textbf{Proof of Lemma \ref{lem:grad1}:} As described in Appendix \ref{app:singleparam}, let $\hat{\p}$ be a set of near optimal posted prices used by an OPM, which can be found according to Theorem \ref{thm:k_uniform_rev}. We define the prices of our sequential public model in a way that ensures that for every agent $i$ and for every $j< k$, $T_i^j=\hat{p}_i$ (and $T_i^j=\infty$ for $j\geq k$). 
This can be done using the following:
\begin{itemize}
	\item For agent $n$, set $p_n^j=\hat{p}_n\cdot\left(1-w_j\right)$ for every $j< k$, and $p_n^j=\infty$  for $j\geq k$. Set $r_{n,j}^j=F_n(T_n^j)$, $r_{n,j}^{j+1}=1-F_n(T_n^j)$. Notice this enforces the thresholds we described above, and the $r_{n,j}^k$'s are matching our definitions.
	\item For agent $i<n$, given $r_{i+1,j}^k$ for every $j$ and $k$, we set $p_i^j= \hat{p}_i (1-\sum_{\ell\geq j} w_\ell\cdot r_{i+1,j}^\ell )$ for every $j< k$, and $p_i^j=\infty$ for $j\geq k$. Set $$r_{i,j}^k=F_i(T_i^j)\cdot r_{i+1,j}^k +(1-F_i(T_i^j))\cdot r_{i+1,j+1}^k.$$ Given that the $r_{i+1,j}^k$'s are computed correctly, this enforces the thresholds we described above, and the  $r_{i,j}^k$'s also match the computed thresholds.
\end{itemize}
Notice that for every $i$ and  $j<\min(i,k)$, we have that:
\begin{eqnarray}
p_i^j & = & \hat{p}_i \left(1-\sum_{\ell\geq j} w_\ell\cdot r_{i+1,j}^\ell\right) \nonumber \\
& \geq & \hat{p}_i \left(1-\sum_{\ell= j}^{k} w_k\cdot r_{i+1,j}^\ell - \sum_{\ell= k+1}^{n} w_\ell \cdot r_{i+1,j}^\ell\right) \nonumber\\
& = & \hat{p}_i \left(1-w_k\sum_{\ell= j}^{k}  r_{i+1,j}^\ell\right)\nonumber\\
& = &  \hat{p}_i (1-w_k),\label{eq:price_grad_lb}
\end{eqnarray}
where the second equality stems from the fact that after selling $k$ items, all prices are set to $\infty$, which implies that $r_{i+1,j}^\ell=0$ for $l>k$, and the last equality is because $\sum_{\ell= j}^{k}  r_{i+1,j}^\ell=1$.

Let $q_{i}^{< k}$ denote the probability that less than $k$ items  were sold prior to the arrival of agent $i$. Setting the thresholds and prices as described above, we get that the expected revenue is:
\begin{eqnarray}
\sum_{i,j} q_{i,j}\cdot p_i^j\cdot (1-F_i(T_i^j)) & = & \sum_{i}\sum_{j=0}^{\min(i-1,k-1)} q_{i,j}\cdot p_i^j\cdot (1-F_i(\hat{p}_i))\nonumber\\
&\stackrel{(\ref{eq:price_grad_lb})}{\geq} & \sum_{i}\sum_{j=0}^{\min(i-1,k-1)} q_{i,j}\cdot \hat{p}_i (1-w_k)\cdot (1-F_i(\hat{p}_i))\nonumber\\
& = & (1-w_k)\sum_{i}q_{i}^{< k} \cdot \hat{p}_i\cdot (1-F_i(\hat{p}_i))\nonumber\\
& \geq & (w_{2k+1}-w_k)\sum_{i}q_{i}^{< k} \cdot \hat{p}_i\cdot (1-F_i(\hat{p}_i))\nonumber\\
& \geq & (w_{2k+1}-w_k)\cdot \Omega(1)\cdot \RR^*(\F,\feas_{k})\nonumber\\
& \geq &  \Omega(1)\cdot (w_{2k+1}-w_k)\cdot\RR^*(\F,\feas_{2k+1}),\label{eq:grad_log}
\end{eqnarray}
where the first equality stems from the fact that we set $T_i^j=\hat{p}_i$ for $j<k$, and the third inequality is because $q_{i}^{< k}$ is exactly the probability that agent $i$ is approached in the posted price sale with no externalities when using $\hat{\p}$ as prices. \qedsymb\\

\noindent \textbf{Proof of Lemma \ref{lem:grad2}:} Let $\hat{\TT}$ be a set of thresholds which approximate $\rbestseq(\F)$ in the full externalities model, and let $\hat{\p}=\tr^{-1}(\hat{\TT})$  be the prices for which $\hat{\TT}$ is the equilibrium in that model. By Theorem \ref{thm:canonical-rev}, the revenue achieved using these prices is $\Theta(1)\cdot \RR^*(\F,\feas_{1})$; \textit{I.e.},
\begin{eqnarray}
\sum_i \left(\prod_{j<i}F_j(\hat{T}_j)\right)\cdot \hat{p}_i\cdot(1-F_i(\hat{T}_i))= \Theta(1)\cdot \RR^*(\F,\feas_{1}).\label{eq:fully_rev_bound}
\end{eqnarray} 

In the model with availability-based externalities, we set prices $\p$ such that for every agent $i$, $T_i^0=\hat{T}_i$ and $T_i^j=\infty$ for every $j>0$; \textit{I.e.}, sell to the agent with probability $F_i(\hat{T}_i)$ \textbf{only if} no agent purchased a good prior to the arrival of the agent. These thresholds can be computed using the following:
\begin{itemize}
	\item For agent $n$, set $p_n^0=\hat{T}_n$ and $p_n^j=\infty$ for $j>0$. 
	\item For agent $i<n$, set: 
	\begin{eqnarray}
	p_i^0  =  \hat{T}_i\cdot(1-\sum_{\ell\geq 0}w_{\ell}\cdot r_{i+1,0}^{\ell})
	=  \hat{T}_i\cdot(1-w_{1}\cdot r_{i+1,0}^{1})
	=  \hat{T}_i\cdot\left(1-w_{1}\cdot \left(1-\prod_{j>i}F_j(\hat{T}_j)\right)\right),\label{eq:threshold_prices_grad}
	\end{eqnarray}
	and $p_i^j=\infty$ for $j>0$.
	%We note that by setting the thresholds as described above, we have that $r_{i+1,0}^1=\left(1- \prod_{j>1}F_j(T_j) \right)$ for 
\end{itemize}
Note that by (\ref{eq:threshold_prices_grad}), we have that $$p_i^0= \hat{T}_i\cdot\left(1-w_{1}\cdot \left(1-\prod_{j>i}F_j(\hat{T}_j)\right)\right)\geq \hat{T}_i\cdot(1- (1-\prod_{j>i}F_j(\hat{T}_j)))=\hat{T}_i\cdot\prod_{j>i}F_j(\hat{T}_j)\stackrel{(\ref{eq:eq_condition_seq})}{=}\hat{p}_i.$$ % where the last equality is due to the equilibrium condition of the full externalities sequential model (\ref{eq:eq_condition_seq}).

We get that the revenue in the model with availability-based externalities is:
\begin{eqnarray*}
	\sum_{i,j}q_{i,j}\cdot p_i^j\cdot(1-F_i(T_i^j)) & = & \sum_{i}q_{i,0} \cdot p_i^0\cdot (1-F_i(T_i^0))\\
	& \geq & \sum_{i}\left(\prod_{j<i}F_j(T_j^0)\right) \hat{p}_i(1-F_i(\hat{T}_i))\\
	& = & \sum_{i}\left(\prod_{j<i}F_j(\hat{T}_j)\right) \hat{p}_i(1-F_i(\hat{T}_i))\\
	& = & \Theta(1)\cdot \RR^*(\F,\feas_{1}).
\end{eqnarray*}\qedsymb
\begin{comment}
\section{Proof of Theorem \ref{thm:grad_main}}\label{app:grad_analysis}
\input{app_grad}
\end{comment}

\section{Irregular distributions}\label{app:irregular}
Some of our results can be extended to hold for irregular distribution (with a constant loss in the approximation). 
In this section we give the key ideas required for these extensions.

\subsection{Simultaneous case}
\paragraph{\textbf{Full externalities:}} In this case, the optimal mechanism for the EAR is no longer a posted price mechanism, but a mechanism that randomizes over prices. Derandomizing the prices will not work, since the mechanism  might sell more than a single item in expectation. 
We observe that our technique of transforming prices as in Theorem \ref{thm:exante_reduction} (i.e., reducing the price of the expensive items and not selling the inexpensive items) works given any posted price mechanism that guarantees a good approximation for any order when selling a single private good. Therefore, the existence of posted prices that obtain half of Myerson's revenue, as guaranteed by the prophet inequalities (see \cite{hartlineMDnA}), yields an approximation factor of $2*5.83$.

\paragraph{\textbf{Status-based externalities:}} The proof of the status-based externalities follows almost identically as in the regular case (Theorem~\ref{thm:semi_sim_main}), as the upper bound in Eq. \eqref{eq:rev_bound_fade_sim} and the proof of Lemma \ref{lm:rev_priv_sim} still hold. The only difference is in the proof of Lemma \ref{lm:rev_pub_sim}, where one cannot use the anonymous prices devised in \cite{DBLP:journals/corr/AlaeiHNPY15}. This is because they are only valid for regular distributions. 
Instead, one may again use the prices that come out of the prophet inequalities, and the price transformation used in Theorem \ref{thm:exante_reduction} to prove Lemma \ref{lm:rev_pub_sim}.

\subsection{Sequential case}
\paragraph{\textbf{Full externalities:}} In this case, we note that the mechanism that arises from the prophet inequalities in the irregular case randomizes over prices. Derandomizing the prices can generate prices $\p$ that guarantee at least half of the optimal revenue, but the item might be sold with probability greater than half (which breaks the previous  proof).
However, for atomless distributions, it is possible to find prices $\hat{\p}$\footnote{This is done by setting $\hat{p}_i=F_i^{-1}(1-\frac{1}{2}\prod_{j<i}\frac{F_j(p_j)}{F_j(\hat{p}_j)}\cdot (1-F_i(p_i))$.} that give a 4-approximation to the optimal revenue, and that the item is sold with probability at most half. Setting prices that induce thresholds as in the proof of Theorem \ref{thm:canonical-rev} generates at least a half of the revenue that can be achieved with $\hat{\p}$, and therefore give a 8-approximation for the irregular case.

\paragraph{\textbf{Status-based externalities:}} The same reduction described in Theorem \ref{thm:semi_seq_main} can be used in the case of irregular distributions. Namely, if there exists a $c$-approximation algorithm for the full externalities model, then there is a $(c+2)$-approximation for the status-based externalities, and therefore a 10-approximation for the irregular case is achievable.

\section{Logarithmic inapproximability using non-discriminatory prices when selling goods with status-based externalities}\label{app:inapprox_homog}
In this section we show that as the $w_i$'s get smaller, posting a single non-discriminatory price for all agents cannot approximate the optimal revenue obtained by using discriminatory prices. Namely, when for all $i$, $w_i=0$, there is a gap of $O(\log n)$ between the revenue attainable when using a single price and the revenue attainable when using multiple prices. 

Consider the setting where agent $i$'s value is drawn from a uniform distribution $F_i=U\left[\frac{1}{i},\frac{1}{i-\frac{1}{2}}\right]$. Since for all $i$, $w_i=0$, the agents face a private sale, and there is no difference between the simultaneous and sequential models. Consider a seller that posts a price $p_i=\frac{1}{i}$ for every agent $i$. The revenue of such a seller is $\RR(\F, \p)= \sum_i p_i \cdot (1-F_i(p_i))= \sum_i \frac{1}{i}\cdot (1-F_i(1/i))\geq \ln n$. 

Consider now a seller that posts a uniform price $p\in \left[\frac{1}{n}, 2\right]$. Assume that the seller posts a price $p\in \left[\frac{1}{j},\frac{1}{j-\frac{1}{2}}\right]$ for some $j\in [n]$. The revenue of such a seller is at most 
\begin{eqnarray*}
\RR(\F, \p)&= & \sum_i p \cdot (1-F_i(p))\leq\frac{1}{j-\frac{1}{2}}\cdot\sum_i(1-F_i(1/j)) \\
&=& \frac{1}{j-\frac{1}{2}}\cdot\sum_{i\leq j}(1-F_i(1/j))= \frac{j}{j-\frac{1}{2}}\leq 2.
\end{eqnarray*}
Therefore, the revenue obtained using discriminatory prices can be greater  by a $\Omega(\log n)$ factor than the revenue using a non-discriminatory price.

\section{An example where the optimal revenue is adaptive} \label{app:example-menu}

We consider the availability based externalities model in sequential sales. When selling goods with social status-based externalities, we show (Theorem \ref{thm:simple_prices_semi}) that a seller who offers two prices per agent, one for the case where no agent bought a good, and one for the case where at least one agent bought a good, is sufficient in attaining optimal revenue. Common wisdom suggest that for the case of availability-based externalities, posting a price vector per agent, which indicates the price offered to the agent for any number of agents who bought a good prior to her arrival, suffices. We show that this is not the case. Namely, we show that for three agents that arrive sequentially and valuations are all drawn uniformly from $[0,1]$, the seller obtains a higher revenue when offering agent $3$ a different price when only agent $1$ bought a good and when only agent $2$ bought a good before she arrived.

%A fully adaptive seller sets a price for every agent $i$ and for every set $S \subseteq 2^{\{1,\ldots,i-1\}}$, which means in the case of three agents to set $7$ prices --- $p_1^\emptyset,p_2^\emptyset,p_2^{\{1\}},p_3^\emptyset,p_3^{\{1\}},p_3^{\{2\}},p_3^{\{1,2\}}$.
We consider the case where the function $w$ which defines the externalities has values $w(0)=0$, $w(1)=0.5$ and $w(2)=0.8$. In this case, we get that the optimal prices are $p_1=0.4360554077$, $p_2(\emptyset)=0.5510244945$, $p_2(\{1\})=0.2272487784$, $p_3(\emptyset)=0.6757323434$, $p_3(\{1\})=0.3119589780$, $p_3(\{2\})=0.3040872295$ and $p_3(\{1,2\})=0.1$ which yields a revenue of $\rev_{1\neq 2} = 0.4622033133$. When restricting the seller to set prices such that $p_3(\{1\})=p_3(\{2\})$, we get that the seller maximizes her revenue by setting prices 
%In the case where the externality function is $w(0)=0,w(1)=0.5,w(2)=0.8,w(3)=1$ the best revenue obtained by a fully adaptive seller is when the prices are set 
to be $p_1=0.4363307363$, $p_2(\emptyset)=0.5511388116$, $p_2(\{1\})=0.2265104177$, $p_3(\emptyset)=0.6758027844$, $p_3(\{1\})=p_3(\{2\})=0.3078610708$ and $p_3(\{1,2\})=0.1$ which yields a revenue of $\rev_{1= 2}=0.4621905314 \ < \ \rev_{1\neq 2}$.
 
%$p_1^\emptyset=0.4360554077,p_2^\emptyset=0.5510244945,p_2^{\{1\}}=0.2272487784,p_3^\emptyset=0.6757323434,p_3^{\{1\}}=0.3119589780,p_3^{\{2\}}=0.3040872295,p_3^{\{1,2\}}=0.1$, and the revenue obtained by these prices is $0.4622033133$.
%While a seller that can only adapt the prices according to how many purchased a good before(i.e. a seller that sets prices $p_1^0,p_2^0,p_2^1,p_3^0,p_3^1,p_3^2$ where $p_i^j$ is the price offered to agent $i$ if $j$ agents purchased a good prior to her arrival) can gain at most $0.4621905314$ by setting the prices to be $p_1^0=0.4363307363,p_2^0=0.5511388116,p_2^1=0.2265104177,p_3^0=0.6758027844,p_3^1=0.3078610708,p_3^2=0.1$.

\end{document}